\documentclass[aps,pre,reprint,groupedaddress,showpacs]{revtex4-1}
\usepackage{graphicx}
\usepackage{color} %eu inclui esta

\begin{document}

%\preprint{}


\title{Horns of subaqueous barchan dunes: {A} study at the grain scale\\
\textnormal{Accepted manuscript for Physical Review E, 100, 042904 (2019), DOI: 10.1103/PhysRevE.100.042904}}



\author{Carlos A. Alvarez}
 \email{calvarez@fem.unicamp.br}
\author{Erick M. Franklin}
 \email{franklin@fem.unicamp.br}
 \thanks{Corresponding author}
\affiliation{%
School of Mechanical Engineering, UNICAMP - University of Campinas,\\
Rua Mendeleyev, 200, Campinas, SP, Brazil\\
 % \textbackslash\textbackslash
}%
%\thanks{}


\date{\today}

\begin{abstract}

Many complex aspects are involved in the morphodynamics of crescent-shaped dunes, known as barchans. One of them concerns the trajectories of individual grains over the dune, and how they affect its shape. In the case of subaqueous barchans, we proposed in Alvarez and Franklin [Phys. Rev. Lett. 121, 164503 (2018)] that their extremities, called horns, are formed mainly by grains migrating from upstream regions of the initial pile, and that they exhibit significant transverse displacements. Here, we extend our previous work to address the dynamics of grains migrating to horns after the dune has reached its crescentic shape, and present new aspects of the problem. In our experiments, single barchans evolve, under the action of a water turbulent flow, from heaps of conical shape formed from glass beads poured on the bottom wall of a rectangular channel. Both for evolving and developed barchans, the horns are fed up with grains coming from upstream regions of the bedform and traveling with significant transverse components, differently from the dynamics usually described for the aeolian case. For these grains, irrespective of their size and strength of water flow, the distributions of transverse and streamwise components of velocities are well described by exponential functions, with the probability density functions of their magnitudes being similar to results obtained from previous studies on flat beds. Focusing on moving grains whose initial positions were on the horns, we show that their residence time and traveled distance are related following a quasi-linear relation. Our results provide new insights into the physical mechanisms underlying the shape of barchan dunes.

\end{abstract}

%\pacs{45.70.Qj, 92.40.Pb}
%\keywords{}

\maketitle

%\section{Introduction}

\section{\label{sec:Intro} INTRODUCTION}

\begin{figure}[b]
\includegraphics[width=0.6\linewidth]{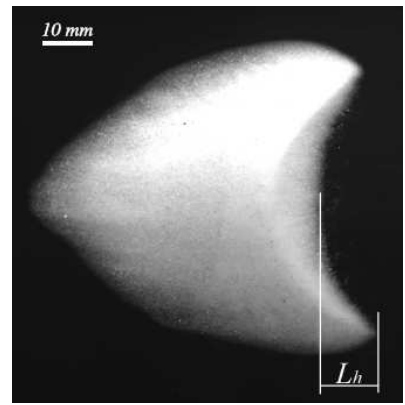}
\caption{Top view of a subaqueous barchan dune with its horn length $L_h$ displayed. In this figure, flow is from left to right.}
	\label{fig:barchan_top_view}
\end{figure}

Sand dunes are the result of complex physical interactions between a fluid flow and an extended area of sand \cite{Bagnold_1, Hersen_1}, their shape and dynamics depending on certain conditions such as the direction and strength of the flow and the available amount of sand. Under one-directional fluid flow with moderate shear stresses, so that bed load is the main mode of sand transport, barchans grow \cite{Bagnold_1, Herrmann_Sauermann, Hersen_3}. Barchan dunes are characterized by their crescentic shape with horns pointing downstream (Fig. \ref{fig:barchan_top_view}), and are frequently found in both nature and industry, some examples being barchans found in deserts, rivers, petroleum pipelines, and even on the surface of Mars \cite{Claudin_Andreotti, Parteli2}. Because of their robust shape in considerably different scales and the large number of environments where they are found, barchan dunes have been studied over the last century by a great number of scientists attracted by the problem. To list but a few examples, Refs. \cite{Bagnold_1, Herrmann_Sauermann, Sauermann_4, Hersen_1, Andreotti_1, Andreotti_2, Kroy_A, Kroy_C, Kroy_B, Groh1, Franklin_8, Lammel, Parteli4, Khosronejad, Alvarez, Alvarez3, Wang_Ch, Gadal} investigated the problem analytically, experimentally or numerically. However, given the high complexity of grain-fluid interactions and the different scales involved, the problem is still open and several aspects need to be understood before a complete understanding is achieved.

Previous analytical studies, based on mechanistic approaches and stability analyses, increased our understanding on the growth and dynamics of barchan dunes, explaining, for example, the dependence of the barchan velocity on the inverse of its size \cite{Bagnold_1} and that the fluid flow is the unstable mechanism for the growth of dunes \cite{Engelund_1, Engelund_Fredsoe}. Experimental data showed, among other things, that, indeed, these conclusions are true \cite{Hersen_1, Andreotti_1, Andreotti_2, Elbelrhiti}. However, despite the large number of experimental works available, only a few of them presented measurements at the grain scale \cite{Alvarez3}. Numerical works, for a great part, used information from analytical and experimental studies to model granular matter as a continuum medium in order to allow the simulation of real dunes consisting of a large number of grains \cite{Sauermann_4, Herrmann_Sauermann, Kroy_A, Kroy_C, Kroy_B, Schwammle, Parteli4}. In those continuum models, given the lack of experimental measurements at the grain scale, the flow of grains is supposed to be mainly longitudinal, with some lateral diffusion. Although some of the recent numerical studies simulate the grains as a discrete medium by using, for example, the discrete element method (DEM) \cite{Kidanemariam}, continuum models are still important to simulate large barchan fields for which the number of grains does not allow discrete simulations at present. In continuum models, information such as typical trajectories and characteristic lengths and times are essential to fit adjustable constants. In the case of simulations employing DEM, they are computationally expensive and comparisons with experiments are still necessary to validate their results.

Concerning the horns, there is a small number of studies devoted to their growth and stability. Khosronejad and Sotiropoulos \cite{Khosronejad} presented a numerical investigation on the instabilities on the surface of horns of existing barchans. They found that transverse waves propagating over the horns give rise to new barchans, and that their amplitudes and wavelengths are related. Hersen \cite{Hersen_3} and Schw{\"a}mmle and Herrmann \cite{Schwammle} investigated numerically the formation of aeolian barchans from different initial shapes using continuum models, in which lateral diffusion was included to account for part of the transverse displacements of grains. According to Hersen \cite{Hersen_3}, the physical origin of the lateral diffusion included in those models is reptation caused by the impact of salting grains. In particular, Hersen \cite{Hersen_3} proposed that aeolian barchans can be modeled as longitudinal 2D slices that exchange mass among them mainly by lateral diffusion, but also air entrainment and slope effects. Within this picture, the celerity of each slice varying with the inverse of its size \cite{Bagnold_1, Kroy_C, Andreotti_1}, horns grow mainly with grains originally in the lateral flanks of the initial heap. Although this description is generally accepted for aeolian dunes, it has never been experimentally verified in the aeolian case. In the subaqueous case, Alvarez and Franklin \cite{Alvarez3} showed a different picture, described in the next paragraph. In a recent paper \cite{Alvarez}, we investigated the formation of subaqueous barchans from initially conical heaps by considering the growth of horns. We showed that horns grow from an initial instant, given by $0.5 t_c$, until they reach a final length at $2.5 t_c$, where $t_c$ is a characteristic time for the displacement of barchans, computed as the length of the bedform divided by its celerity.

Few experimental measurements at the grain scale are reported for subaqueous bed load. For plane granular beds, Seizilles et al. \cite{Seizilles} investigated bed load under laminar flows and Lajeunesse et al. \cite{Lajeunesse} and Penteado and Franklin \cite{Penteado} investigated the turbulent case. As common results, these works showed that the displacements of individual grains are intermittent and that distributions of grain velocities can be adjusted by exponential functions. Lajeunesse et al. \cite{Lajeunesse} and Penteado and Franklin \cite{Penteado} showed that the streamwise velocities scale with the excess of shear stress, and that transverse velocities are distributed around a zero mean value. Seizilles et al. \cite{Seizilles} proposed that transverse displacements on flat beds are caused by a Fickian diffusion mechanism with a characteristic length of 0.030$d$, where $d$ is the mean grain diameter. In a recent paper \cite{Alvarez3}, we investigated experimentally bed load during the growth of barchans from initial piles of conical shape. We measured the trajectories of grains migrating to the growing horns and showed that most of them came from upstream regions on the periphery of the initial pile, with transverse displacements by rolling and sliding that were considerable. These evidences diverge from the general description for aeolian dunes, where transverse displacements are due mainly to the diffusive effect of reptons and, therefore, horns are expected to grow mainly with grains originally in the lateral flanks of the initially conical pile.

\subsection{\label{sec:Prior work} Prior work}

In previous studies \cite{Alvarez, Alvarez3}, we reported experimental results on the growth of subaqueous barchans under controlled conditions. For each test run, we poured glass beads in a closed conduit to form a conical pile that, afterward, under the action of a turbulent water flow, evolved to a single barchan dune. In the particular case of Alvarez and Franklin \cite{Alvarez3}, bed load was measured at the grain scale and we showed that most of grains going to horns were initially on the periphery of the initial pile and exhibited significant transverse displacements. In that work, the reported data were limited to 0.40 mm $\leq$ $d$ $\leq$ 0.60 mm glass beads and to trajectories measured during the growth of horns.

\subsection{\label{sec:Current work} This study}      

In this paper, we present a thorough investigation on the trajectories of grains going to horns of subaqueous barchans, not only extending the experiments of Alvarez and Franklin \cite{Alvarez3} to other cases, but also presenting new aspects of the problem. In the present experiments, subaqueous dunes were formed from initial piles of conical shape, and we measured the trajectories of grains going to horns for both the cases of evolving dunes (growing horns), $0.5 t_c \,\leq\, t \,\leq\, 2.5 t_c$, and developed dunes (stable horns), $t \,>\,2.5 t_c$. Within the range of parameters of the present study, we show that the grain dynamics previously observed for evolving barchans does not change with the grain size or flow strength, and that the same dynamics is observed for developed barchans. The present results show that the general assertions for aeolian dunes that horns grow and are sustained mainly with grains originally in the transverse extremities of the bedform do not apply for subaqueous barchans. Furthermore, we show that the distributions of transverse and streamwise components of grain velocities are well described by exponential functions, and we find the typical residence time and traveled distance of moving grains whose initial positions were on the horns. Our results change the way in which horns formation and stability and bed load are explained for subaqueous barchans.

In the following, Sec. \ref{sec:Exp} describes the experimental setup and employed methods, Sec. \ref{sec:Res} presents the results for the trajectories of grains going to horns of evolving and developed barchans, and the residence time and traveled distance of grains whose initial positions were on the horns, and Sec. \ref{sec:Conclu} presents the conclusions.

\section{\label{sec:Exp} EXPERIMENTAL SETUP}
\begin{figure*}
\includegraphics[width=0.7\linewidth]{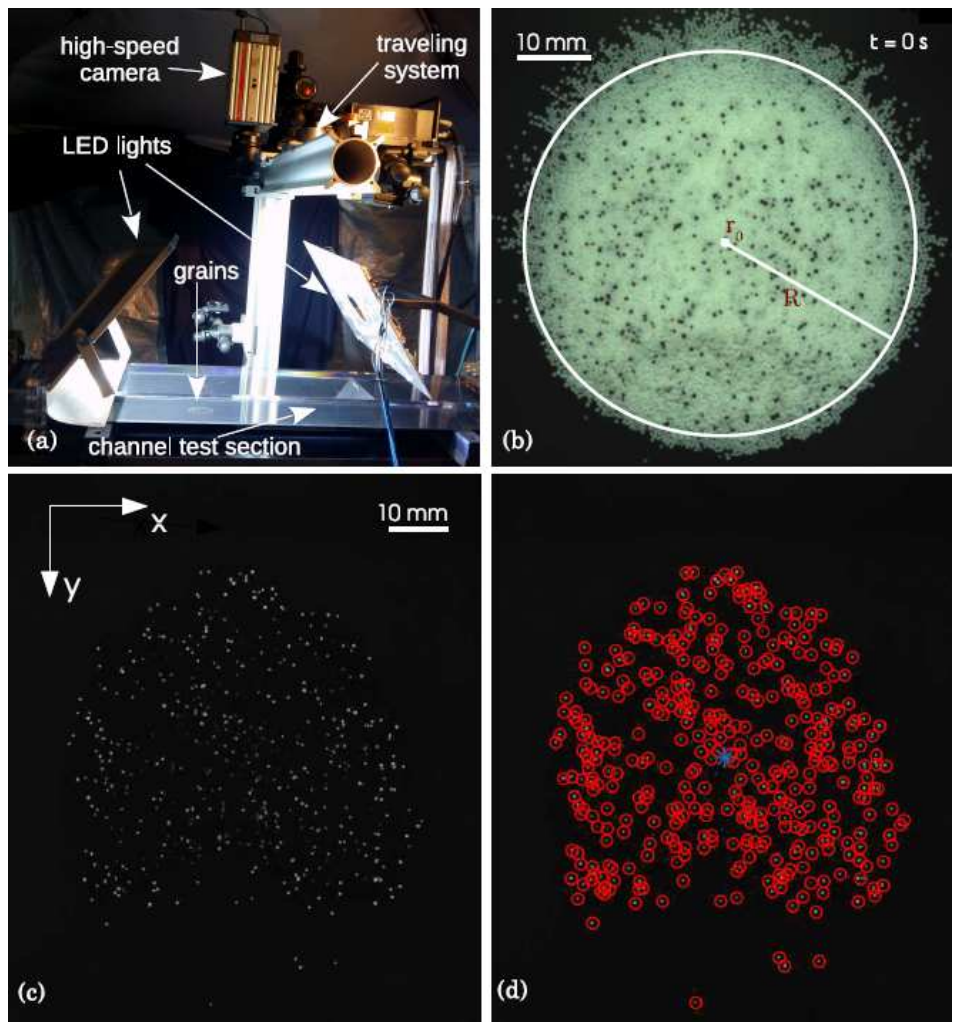}
\caption{Experimental setup, definition of geometrical parameters, and particle detection. (a) Photograph of the experimental setup showing the test section, high-speed camera, traveling system, LED lights, and grains placed on the bottom wall of the channel. (b) Top view of an initially conical heap of radius $R$ at time $t$= 0 s, where $r_0$ is the initial position of the pile centroid and black spots are tracers. (c) Top-view image of a dune in grayscale showing the tracers. In this figure, the flow is from top to bottom. (d) Example of treated image showing the detected particles. In this figure, the red circles are surrounding the tracers identified in (c), and the asterisk corresponds to the instantaneous position of the dune centroid. $Re$ = 1.82 $\times$ 10$^4$ and the heap initial mass was 6.2 g.}
	\label{fig:dune_track}
\end{figure*}

We used the same experimental device as in Refs. \cite{Alvarez, Alvarez3}, which consisted basically of a water reservoir, centrifugal pumps, a 5-m-long closed-conduit channel, a settling tank, and a return line. The channel, made of transparent material, had a rectangular cross section 160 mm wide by 50 mm high (2$\delta$ = 50 mm), and the test section started at 3 m (40 hydraulic diameters) downstream of the channel entrance. Prior to each test, with the channel previously filled with water, controlled grains were poured in the test section, forming a single conical pile at the bottom wall. Afterward, for each test run, a turbulent water flow was imposed, deforming the conical pile into a barchan dune. With this procedure, each experiment concerned one single barchan that loosed grains by its horns, decreasing slowly in size while migrating. Figure \ref{fig:dune_track}(a) presents a photograph of the experimental setup displaying, among other elements, the test section and the initial pile. Figure \ref{fig:dune_track}(b) shows a top view of an initially conical heap, where $R$ is the radius of the initial pile, defined as the maximum radius with the origin at the centroid and that do not contain void regions, and $r_0$ is the initial position of the pile centroid.

The tests were performed with tap water at temperatures between 24 and 26 $^o$C and round glass beads (density of grains $\rho_s$ = 2500 kg/m$^3$ and bulk density of 1500 kg/m$^3$) with 0.15 mm $\leq\,d\,\leq$ 0.25 mm and 0.40 mm $\leq\,d\,\leq$ 0.60 mm. In order to facilitate the tracking of moving grains, 2$\%$ of them were tracers (grains of different color but the same density, diameter and surface characteristics as the other grains). The cross-sectional mean velocities of water $U$ were 0.243, 0.294 and 0.364 m/s, corresponding to Reynolds numbers based on the channel height $Re$ = $\rho U 2\delta /\mu$ of $1.21 \times 10^4$, $1.47 \times 10^4 $ and $1.82 \times 10^4 $, respectively, where $\rho$ is the density and $\mu$ the dynamic viscosity of the fluid. The shear velocities on the channel walls $u_*$ were computed from the velocity profiles measured with a two-dimensional particle image velocimetry device (2D-PIV), and were found to follow the Blasius correlation \cite{Schlichting_1}. By using the hydraulic diameter of the channel, they correspond to 0.0141, 0.0168 and 0.0202 m/s for the three flow rates employed. The initial heaps were formed with 6.2 and 10.3 g of glass beads, corresponding to initial volumes of 4.1 and 6.9 cm$^3$, and to $R$ of 2.6 and 3.2 cm, respectively.

Table. \ref{tab1} summarizes the tested conditions, for which we performed between three and five independent runs for each experimental condition. In Tab. \ref{tab1}, the column \textit{dune condition} refers to whether the measurements concern an evolving or a developed barchan. The table presents also the mass of the initial pile $m_0$, the grain diameter $d$, the channel Reynolds number $Re$, the Reynolds number at the grain scale  $Re_*$ = $\rho u_* d / \mu$, and the Shields number $\theta = (\rho u_*^2)/((\rho_s - \rho )gd)$, which is the ratio between the drag on grains and their relative weight. In the Shields number, $g$ is the magnitude of gravity.

\begin{table}[!ht]	
	\begin{center}
		\begin{tabular}{|c|c|c|c|c|c|c|c|}
			\hline
			Case & $d$ & dune condition & $Re$ & $Re_*$ & $\theta$ & $m_0$ \\ 
			$\cdots$  & (mm) & $\cdots$ & $\cdots$ & $\cdots$ & $\cdots$ & (g) \\\hline
			a & 0.15 -- 0.25 & evolving & 1.21 $\times$ 10$^4$ & 3 & 0.07 & 6.2 \\\hline
			b   & 0.15 -- 0.25 & evolving & 1.47 $\times$ 10$^4$ & 3 & 0.10 & 6.2 \\\hline
			c & 0.15 -- 0.25 & evolving& 1.82 $\times$ 10$^4$ & 4 & 0.14 & 6.2 \\\hline
			d & 0.15 -- 0.25 & developed & 1.21 $\times$ 10$^4$ & 3 & 0.07 & 6.2 \\\hline
			e   & 0.15 -- 0.25 & developed & 1.47 $\times$ 10$^4$ & 3 & 0.10 & 6.2 \\\hline
			f & 0.15 -- 0.25 & developed & 1.82 $\times$ 10$^4$ & 4 & 0.14 & 6.2 \\\hline
			g   & 0.40 -- 0.60 & developed & 1.21 $\times$ 10$^4$ & 7 & 0.03 & 6.2 \\\hline
			h   & 0.40 -- 0.60 & developed & 1.47 $\times$ 10$^4$ & 8 & 0.04 & 6.2 \\\hline
			i   & 0.40 -- 0.60 & developed & 1.82 $\times$ 10$^4$ & 10 & 0.06 & 6.2 \\\hline
			j   & 0.40 -- 0.60 & developed &1.21 $\times$ 10$^4$ & 7 & 0.03 & 10.3 \\\hline
			k   & 0.40 -- 0.60 & developed & 1.47 $\times$ 10$^4$ & 8 & 0.04 & 10.3 \\\hline
			l  & 0.40 -- 0.60 & developed&  1.82 $\times$ 10$^4$ & 10 & 0.06 & 10.3 \\\hline		
		\end{tabular}
	\end{center}
	\caption{Label of tested cases, dune diameter $d$, dune condition during measurements, channel Reynolds number $Re$, Reynolds number at the grain scale $Re_*$, Shields number $\theta$, and mass of the initial heap $m_0$.}
	\label{tab1}
\end{table}

A high-speed camera of complementary metal-oxide-semiconductor type (CMOS) was placed above the channel to record the bed evolution (Fig. \ref{fig:dune_track}(a)). We used a camera with a spatial resolution of 1280 px $\times$ 1024 px at frequencies up to 1000 Hz, controlled by a computer. In our tests, we set the frequency to values within 50 and 200 Hz, depending on the average velocity of grains, and we used a lens of $60$ mm focal distance and F2.8 maximum aperture. Lamps of light-emitting diode (LED) were branched to a continuous current source in order to supply the required light while avoiding beating between the camera and light frequencies. Prior to the beginning of tests, a calibration procedure which consisted of taking one picture from a scale placed in the channel (filled with water) was performed, allowing the conversion from pixels to a physical system of units. We set the region of interest (ROI) to 800 px $\times$ 1024 px to better fit a field of view of 80.0 mm $\times$ 102.4 mm; therefore, the area covered by each grain in the acquired images varied within 2 to 28 px. Examples of movies, showing the motion of grains over evolving and developed barchans, are available as Supplemental Material \cite{Supplemental}.

Once the images were obtained, the centroids of tracers and those of barchans were identified with an image processing code written in the course of this work based on Refs. \cite{Alvarez2, Kelley}. To compute the trajectories of tracers, the code uses a particle tracking velocimetry approach (PTV) that follows each centroid along time.

\section{\label{sec:Res} RESULTS AND DISCUSSION}

\subsection{\label{subsec:Res_A} Pathlines of migrating grains} 

\begin{figure}
	\begin{minipage}[c]{\columnwidth}
		\begin{center}
			\includegraphics[width=.65\linewidth]{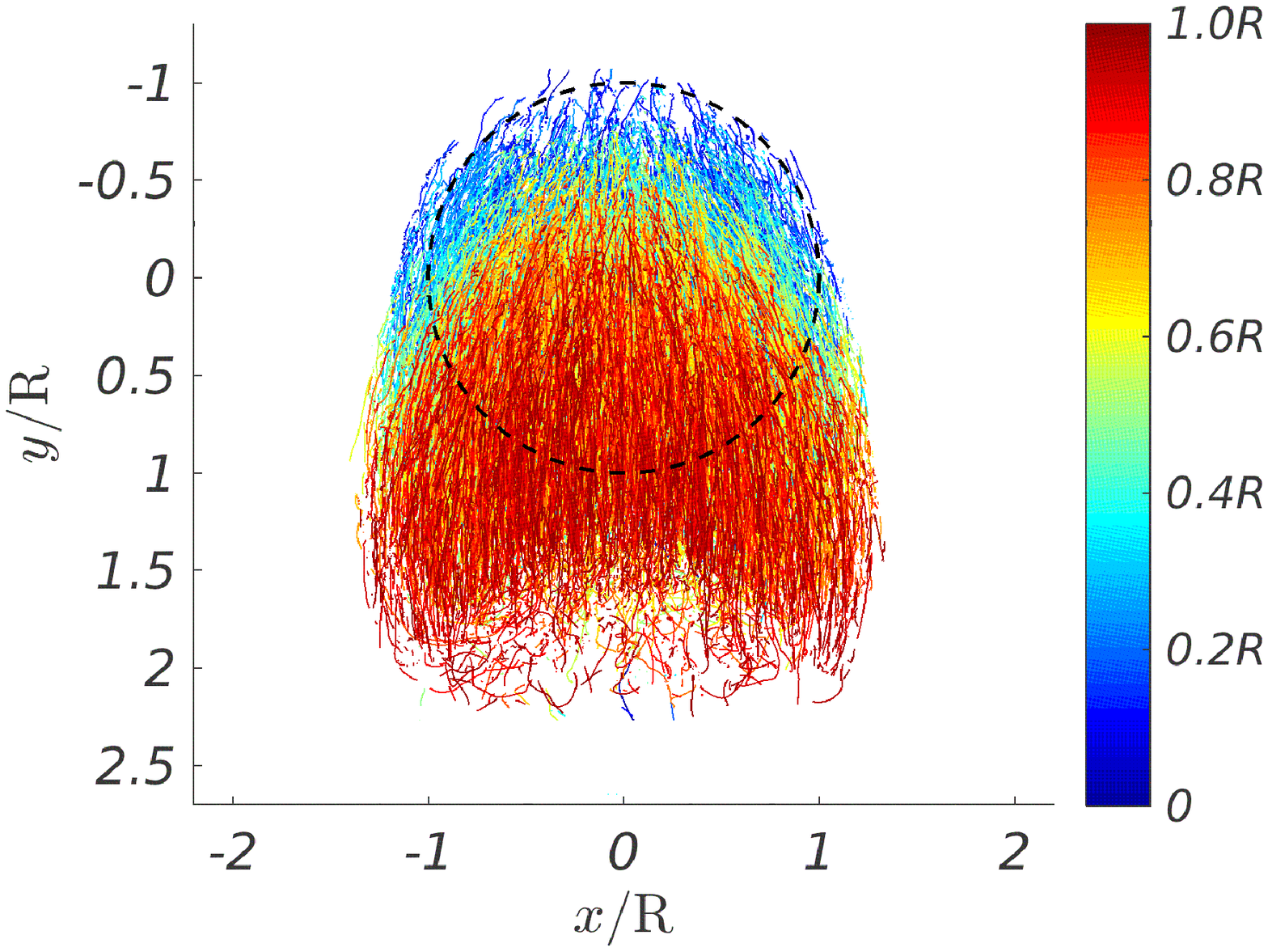}\\
			(a)
		\end{center}
	\end{minipage} \hfill
	\begin{minipage}[c]{\columnwidth}
		\begin{center}
			\includegraphics[width=.65\linewidth]{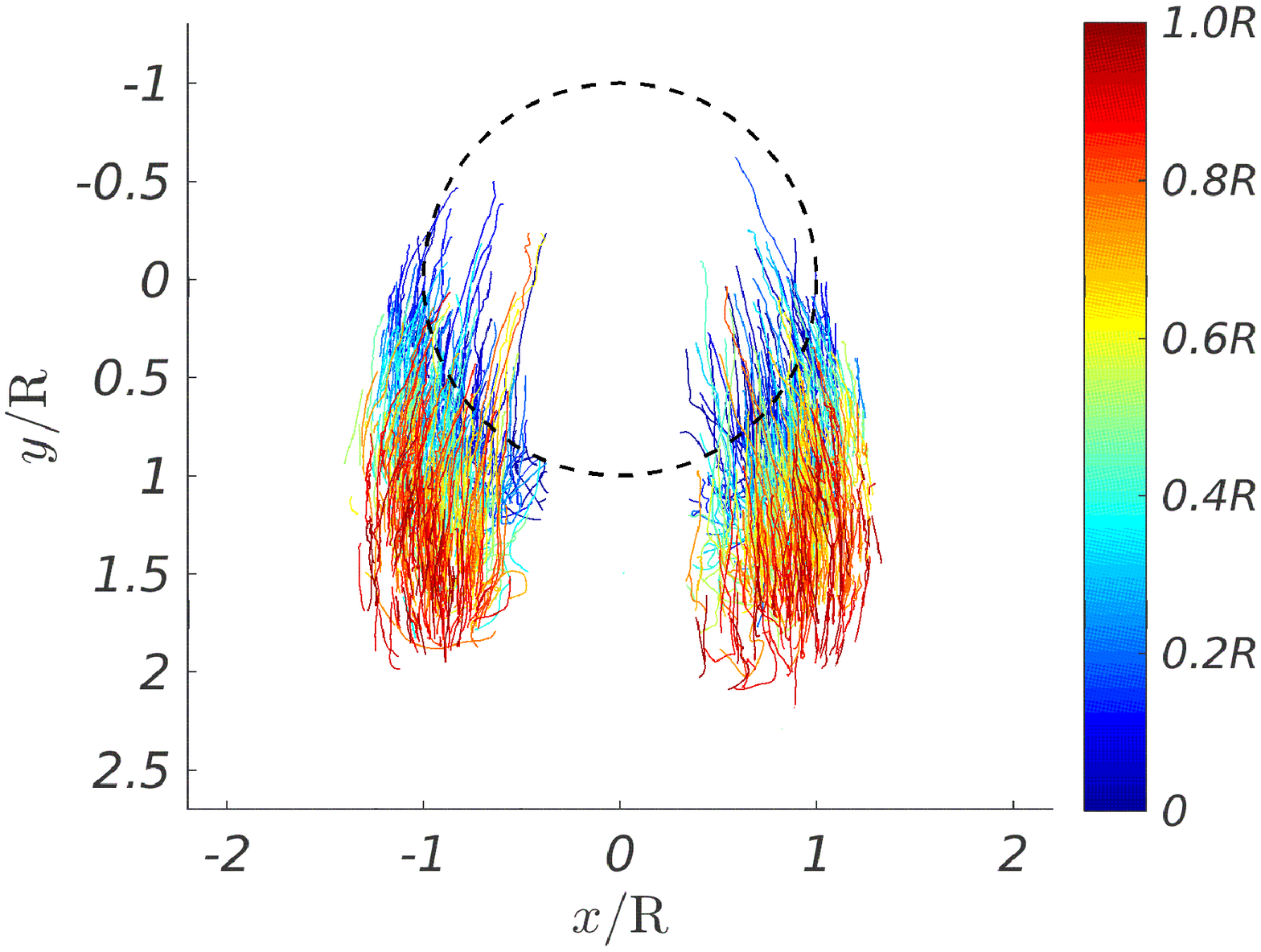}\\
			(b)
		\end{center}
	\end{minipage}
	
\caption{Pathlines of tracers over a evolving dune made of 0.15 mm $\leq\,d\,\leq$ 0.25 mm glass beads. a) All moving tracers during the growth of a barchan dune. b) Tracers that migrated to horns during the growth the barchan dune. The black dashed circle represents the initial pile of radius $R$, and the scaling bar shows the values of $r_c - r_0$ normalized by $R$. The water flow is from top to bottom, $Re$ = $1.82 \times 10^4 $, $m_0$ = 6.2 g, and $R$ = 2.6 cm.}
\label{fig:init_sm_grain}
\end{figure}

In Alvarez and Franklin \cite{Alvarez3} we presented the trajectories of grains migrating to the horns of barchans during their growth from conical piles that consisted of 0.40 mm $\leq\,d\,\leq$ 0.60 mm glass beads. Therefore, in the case of evolving barchans, we present here only the trajectories for 0.15 mm $\leq\,d\,\leq$ 0.25 mm glass beads, the trajectories of larger grains being found in Ref. \cite{Alvarez3}. Figure \ref{fig:init_sm_grain}(a) shows the pathlines of moving tracers during the growth of a barchan dune from an initial conical pile consisting of 0.15 mm $\leq\,d\,\leq$ 0.25 mm glass beads. The initial mass of the pile, $m_0$, was 6.2 g, which corresponded to $R$ = 2.6 cm. The abscissa and ordinate correspond, respectively, to the transverse and streamwise coordinates, $x$ and $y$, normalized by $R$. In Fig. \ref{fig:init_sm_grain}, the black dashed circle displays the initial pile and the color of pathlines varies according to the position of the pile centroid, which moves while the barchan grows. In this way, the blue (upper) and red (lower) pathlines correspond to the initial and final positions of the pile centroid, respectively, the scaling bar showing the values of $r_c - r_0$ normalized by $R$, where $r_c$ is the instantaneous position of the pile centroid and $r_0$ the initial one (see Supplemental Material \cite{Supplemental} for some trajectories superposed with a photograph of the barchan).

Figure \ref{fig:init_sm_grain}(b) shows the pathlines of the tracers that migrated to horns during their growth (see Supplemental Material \cite{Supplemental} for trajectories of other test runs concerning the 0.15 mm $\leq\,d\,\leq$ 0.25 mm beads).  As in the cases presented in Ref. \cite{Alvarez3}, we note that grains experiment significant transverse displacements, many of them describing circular paths while migrating toward the horns.

\begin{figure}
	\begin{minipage}[c]{\columnwidth}
		\begin{center}
			\includegraphics[width=.65\linewidth]{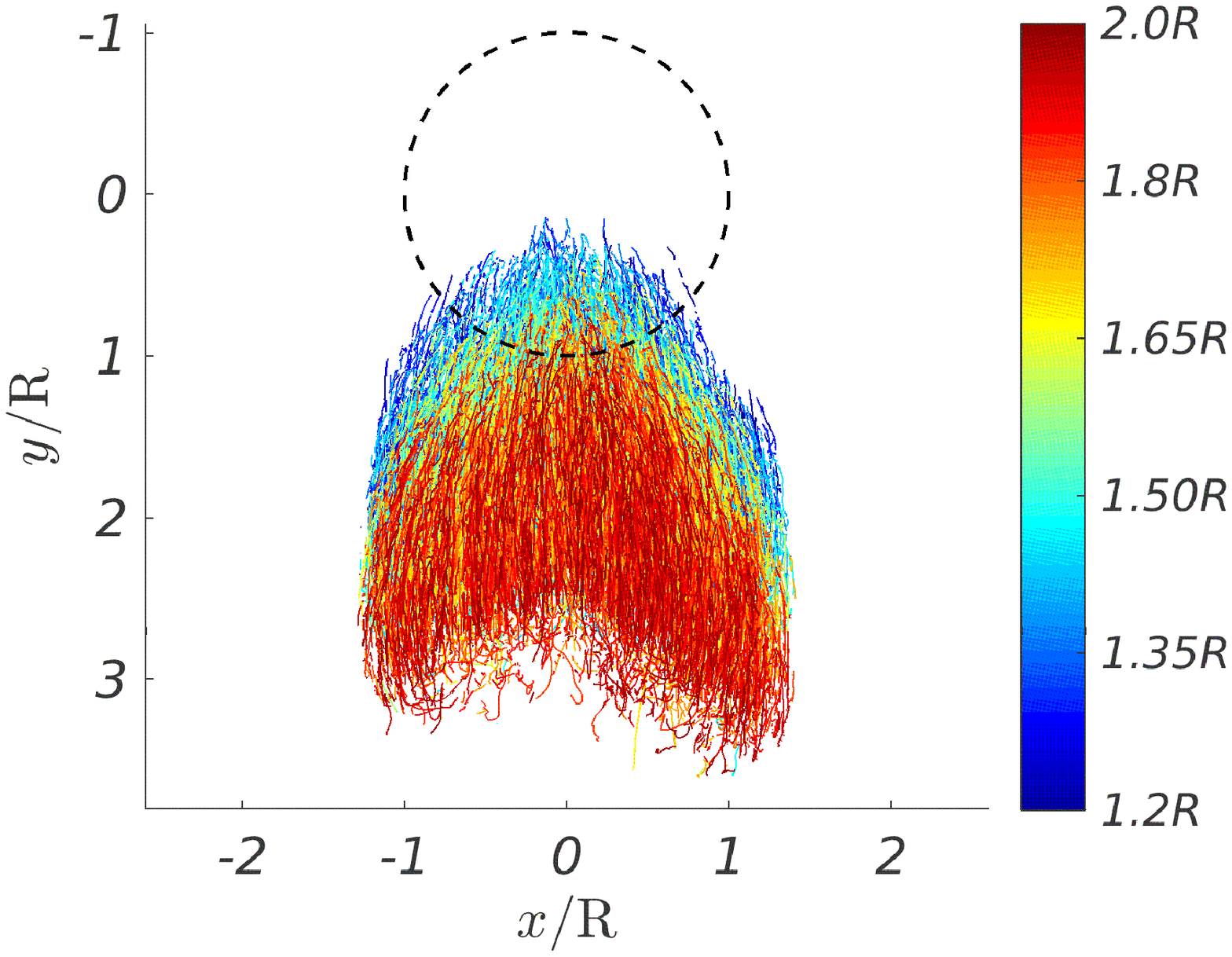}\\
			(a)
		\end{center}
	\end{minipage} \hfill
	\begin{minipage}[c]{\columnwidth}
		\begin{center}
			\includegraphics[width=.65\linewidth]{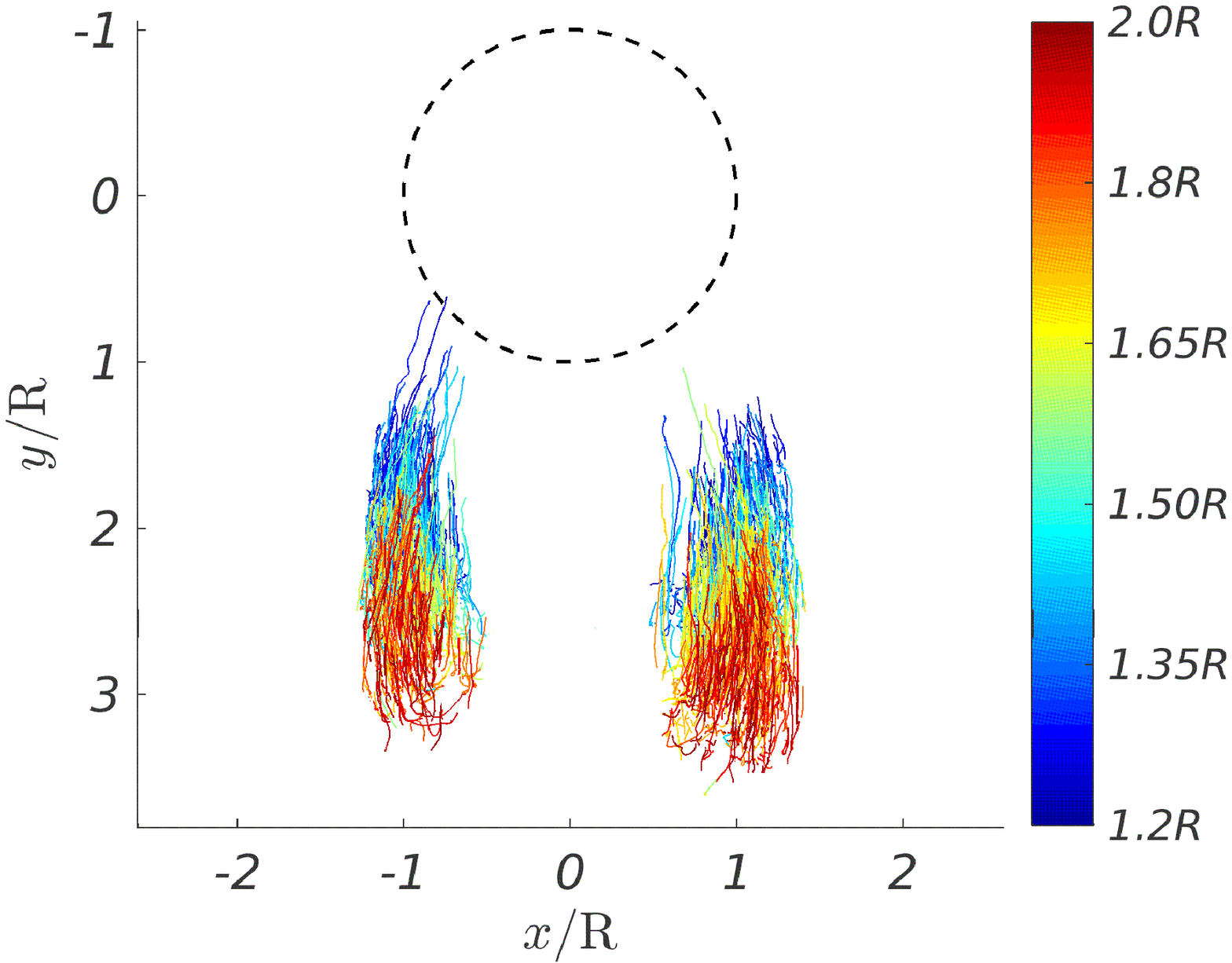}\\
			(b)
		\end{center}
	\end{minipage}
\caption{Pathlines of tracers over a developed dune made of 0.15 mm $\leq\,d\,\leq$ 0.25 mm glass beads. a) All moving tracers over the barchan dune. b) Tracers that migrated to the barchan horns. The black dashed circle represents the initial pile of radius $R$, and the scaling bar shows the values of $r_c - r_0$ normalized by $R$. The water flow is from top to bottom, $Re$ = $1.82 \times 10^4 $, $m_0$ = 6.2 g, and $R$ = 2.6 cm.}
	\label{fig:dev_Sgrain}
\end{figure}

\begin{figure}
	\begin{minipage}[c]{\columnwidth}
		\begin{center}
			\includegraphics[width=.65\linewidth]{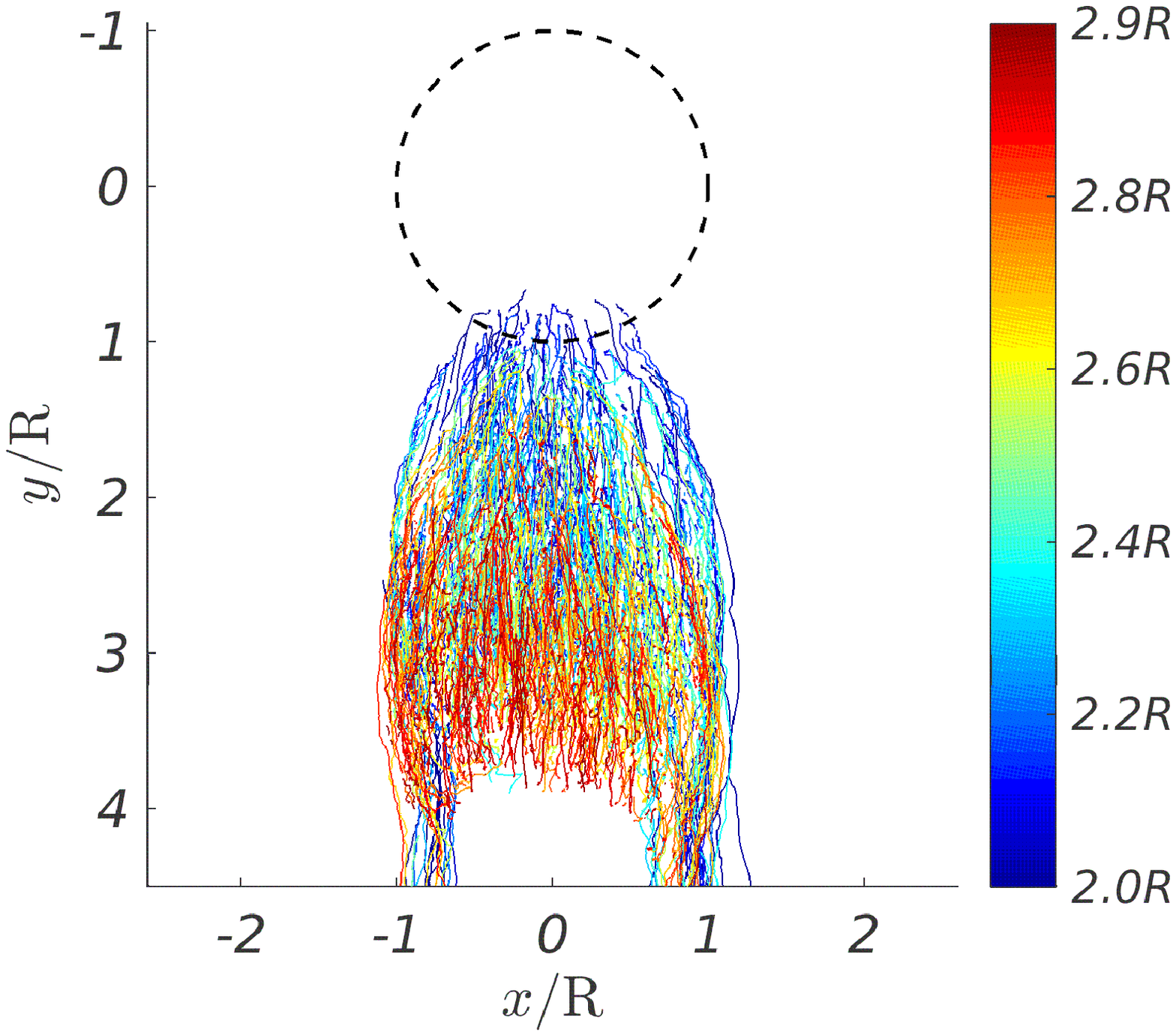}\\
			(a)
		\end{center}
	\end{minipage} \hfill
	\begin{minipage}[c]{\columnwidth}
		\begin{center}
			\includegraphics[width=.65\linewidth]{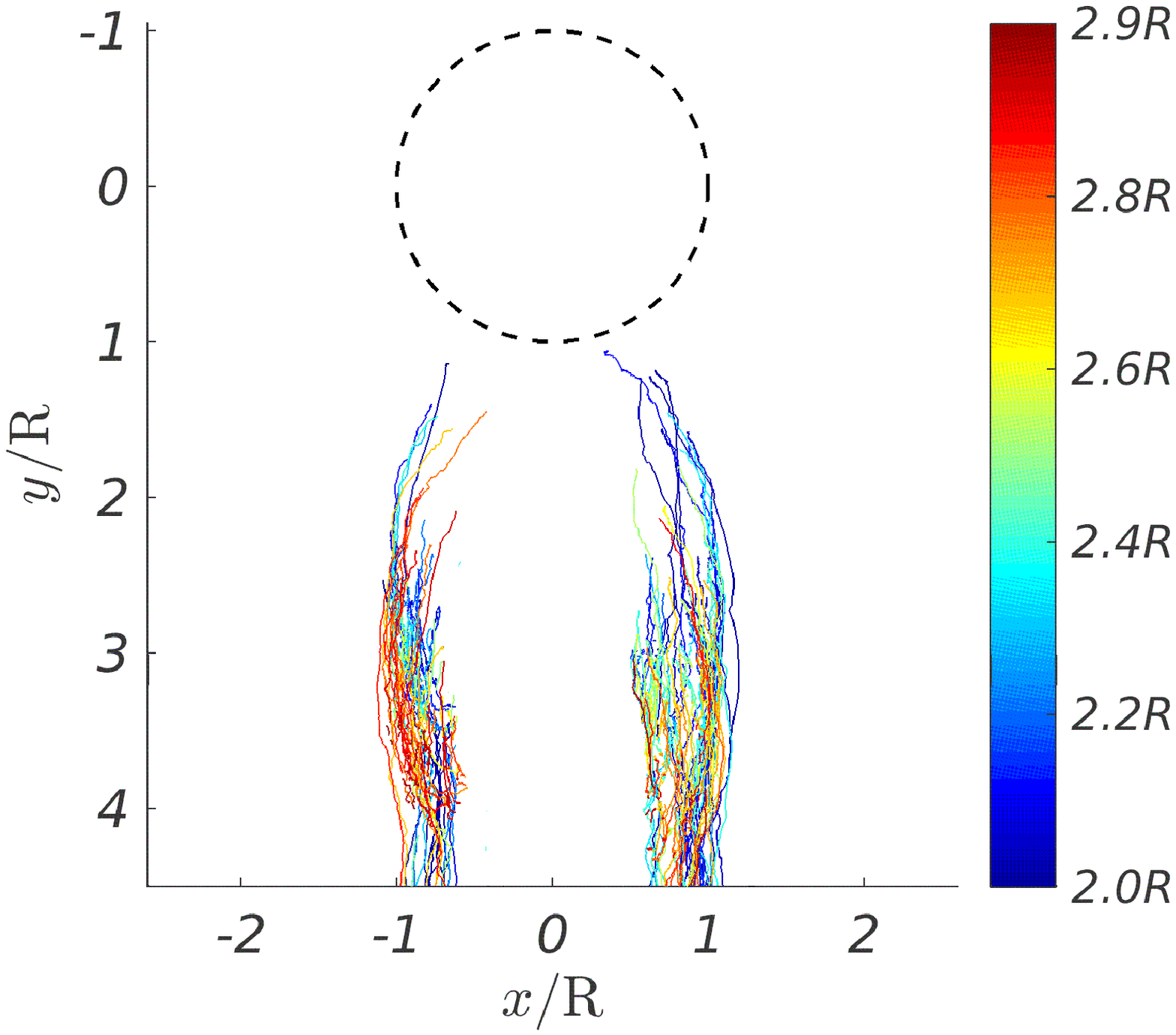}\\	
			(b)
		\end{center}
	\end{minipage}
\caption{Pathlines of tracers over a developed dune made of 0.40 mm $\leq\,d\,\leq$ 0.60 mm glass beads. a) All moving tracers over the barchan dune. b) Tracers that migrated to the barchan horns. The black dashed circle represents the initial pile of radius $R$, and the scaling bar shows the values of $r_c - r_0$ normalized by $R$. The water flow is from top to bottom, $Re$ = $1.47 \times 10^4$, $m_0$ = 6.2 g, and $R$ = 2.6 cm.}
	\label{fig:dev_Bgrain}
\end{figure}

In addition to evolving barchans, we address now the dynamics of grains migrating to horns after the dune has reached its crescentic shape. For that, we identified and tracked the tracers from the moment the dune centroid reached a coordinate 1.0$R$ downstream of its origin, which means that the dune had migrated a distance equivalent to its size and, therefore, all grains within the dune had been displaced. Figure \ref{fig:dev_Sgrain} shows the pathlines of moving tracers over a grown barchan consisting of 0.15 mm $\leq\,d\,\leq$ 0.25 mm glass beads. In this figure, $m_0$ = 6.2 g, which corresponded to $R$ = 2.6 cm, and the colors and legends are the same as in Fig. \ref{fig:init_sm_grain}. Figure \ref{fig:dev_Sgrain}(a) presents the pathlines of all moving tracers, while Fig. \ref{fig:dev_Sgrain}(b) presents the pathlines of traces migrating to the horns. Figure \ref{fig:dev_Bgrain} shows the developed case for $m_0$ = 6.2 g, 0.40 mm $\leq\,d\,\leq$ 0.60 mm and $Re$ = $1.47 \times 10^4$, and the pathlines of the other test runs are available as Supplemental Material \cite{Supplemental}.

From Figs. \ref{fig:dev_Sgrain} and \ref{fig:dev_Bgrain} we observe a behavior similar to that observed for growing barchans, i.e.,  many of moving grains describe circular paths while migrating toward the horns, which implies local transverse components that are significant. These grains come from regions upstream of the dune centroid, moving around the central region of the barchan before reaching the horns. We note also small asymmetries in Figs. \ref{fig:dev_Sgrain} and \ref{fig:dev_Bgrain}, which are due to dispersions in our experiments.

We compare next the mean distance $L_{mean}$ traveled by tracers migrating to horns in the evolving and developed cases. The mean distance was obtained by computing the total distance traveled by each tracer that migrated to the horns, and then taking the arithmetic mean for each tested case. By considering, in addition to cases listed in Tab. \ref{tab1}, those presented in Ref. \cite{Alvarez3}, we found for the evolving barchans

\begin{equation}
14 < L_{mean}/L_{drag} < 30
\end{equation}

\noindent for the 6.2 g piles and

\begin{equation}
18 < L_{mean}/L_{drag} < 22
\end{equation}

\noindent for the 10.3 g piles, while for the developed barchans we found

\begin{equation}
14 < L_{mean}/L_{drag} < 28
\end{equation}

\noindent for the 6.2 g piles and

\begin{equation}
22 < L_{mean}/L_{drag} < 30
\end{equation}

\noindent for the 10.3 g piles, where $L_{drag} = (\rho_s/ \rho)d$ is an inertial length scaling with the length for sand flux saturation \cite{Hersen_1}. The normalized traveled distances are similar for both evolving and developed barchans, corroborating the similar behavior observed from pathlines plotted in Figs. \ref{fig:init_sm_grain} to \ref{fig:dev_Bgrain} and in Refs. \cite{Alvarez3} and \cite{Supplemental}. Their values are three orders of magnitude greater than the diffusion length $\ell_d$ proposed by Seizilles et al. \cite{Seizilles} ($\ell_d/L_{drag} \, \approx$ 0.01), agreeing with values obtained in Ref. \cite{Alvarez3}.

\begin{figure}
	\begin{minipage}[c]{\columnwidth}
		\begin{center}
			\includegraphics[width=.5\linewidth]{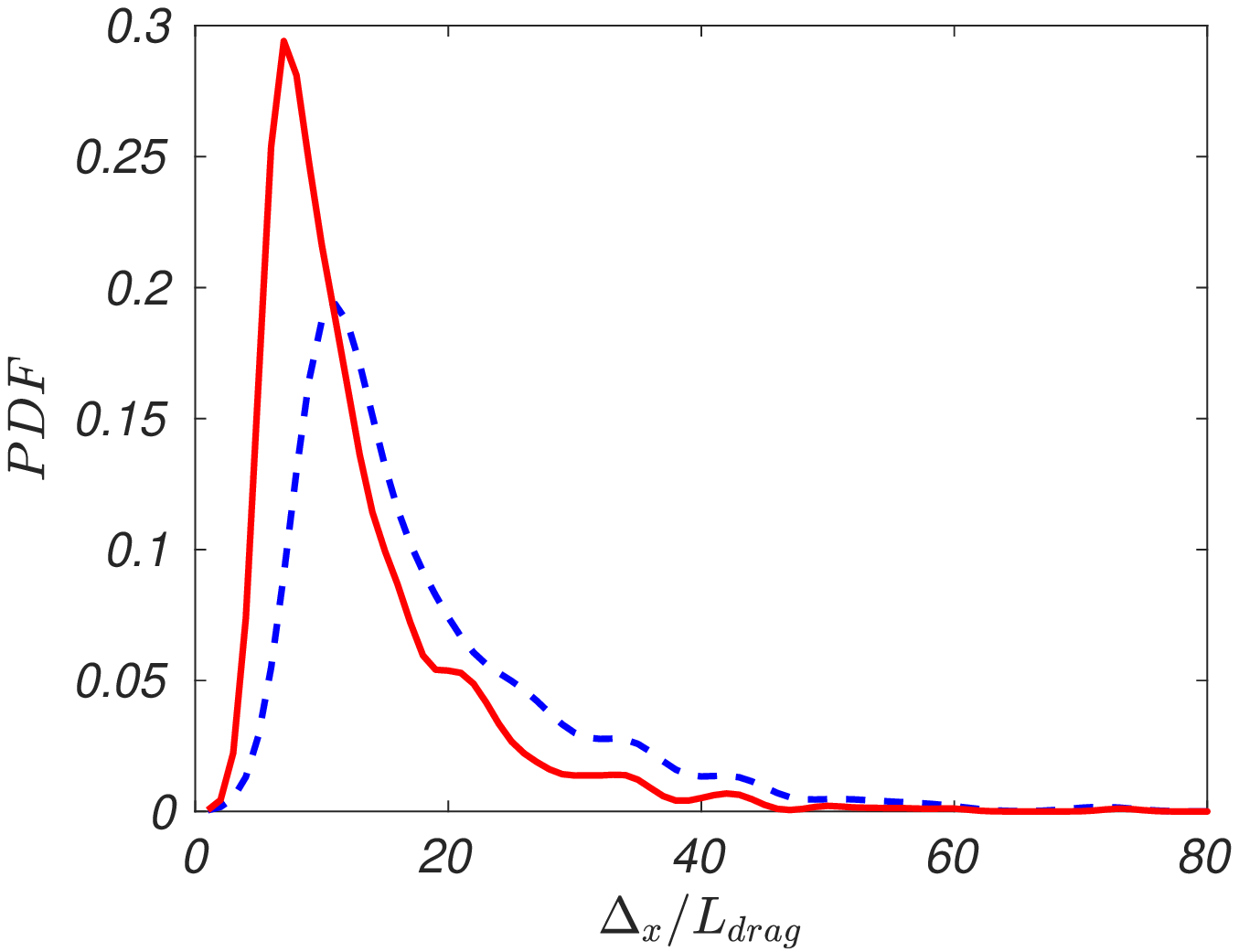}\\
			(a)
		\end{center}
	\end{minipage} \hfill
	\begin{minipage}[c]{\columnwidth}
		\begin{center}
			\includegraphics[width=.5\linewidth]{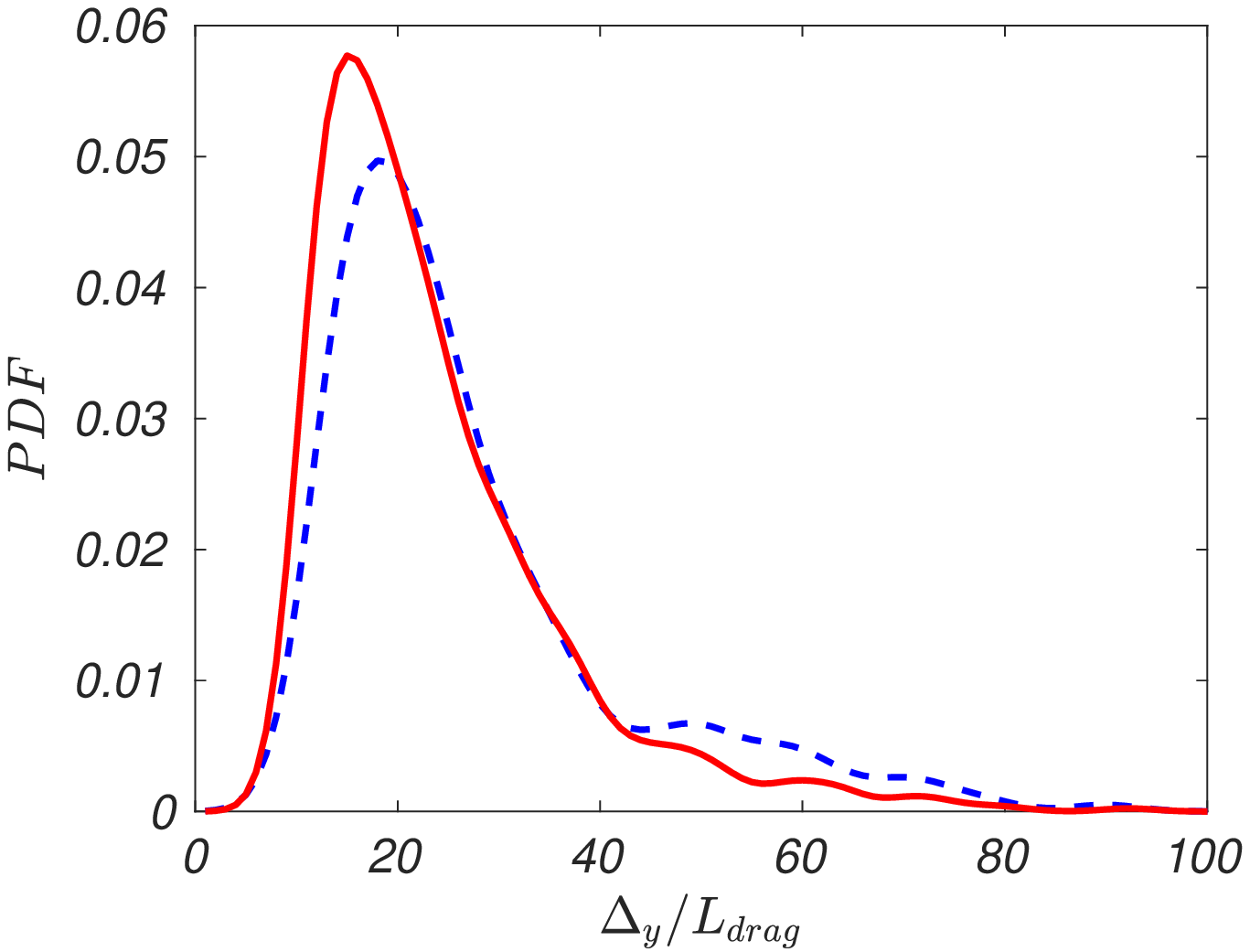}\\	
			(b)
		\end{center}
	\end{minipage}
\caption{PDFs of the total (a) transverse and (b) longitudinal distances, $\Delta_x$ and $\Delta_y$, respectively, normalized by $L_{drag}$ for case k of Tab. \ref{tab1}. Dashed blue lines correspond to the grains that migrated to the horns and continuous red lines to all the other grains.}
	\label{fig:dist_deltas}
\end{figure}

We computed also the total transverse ($\Delta_x$) and longitudinal ($\Delta_y$) distances traveled by the tracers that migrated to horns as well as by the other tracers, both in the evolving and developed cases. Probability density functions (PDFs) can be seen in Figs. \ref{fig:dist_deltas}(a) and \ref{fig:dist_deltas}(b), showing, respectively, $\Delta_x /L_{drag}$ and $\Delta_y /L_{drag}$ for case k of Tab. \ref{tab1}, where the dashed blue lines correspond to the grains that migrated to the horns and continuous red lines to all the other grains (see Supplemental Material \cite{Supplemental} for the same distributions in dimensional form). We observe a significant transverse component in the movement of grains, with higher values for grains migrating to horns in comparison to the others. By considering the most probable values of distributions for the cases listed in Tab. \ref{tab1}, we found 10 $\leq\,\Delta_x /L_{drag}\,\leq$ 12 for the grains migrating to horns, three orders of magnitude greater than the diffusion length $\ell_d$ proposed by Ref. \cite{Seizilles}, and 6 $\leq\,\Delta_x /L_{drag}\,\leq$ 8 for all the other grains. We found also that the ratio between the transverse and longitudinal distances, $\Delta_x / \Delta_y$, is around 0.5 for the grains migrating to horns and 0.4 for all the other grains. These values corroborate the importance of transverse movements in the subaqueous case.

\subsection{\label{subsec:Res_B} Origin of grains migrating to horns}

A suitable method for identifying the locations from where the grains migrate to the horns is the use of radial and angular coordinates with origin at the dune centroid, as presented in Ref. \cite{Alvarez3}. For that, we identified the initial position of each tracer that migrated to the horns in polar coordinates $(|r_1 - r_c|,\,\phi )$, where $r_1$ is the initial position of the tracer, $r_c$ is the instantaneous position of the dune centroid, and $\phi$ is the angle with respect to the transverse direction. With the initial positions, we computed the probability density functions and frequencies of occurrence for all tested conditions.

\begin{figure}[ht]
	\includegraphics[width=0.65\linewidth]{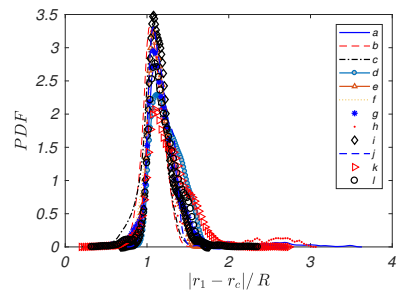}
	\caption{PDFs of the original position of the tracers that migrated to horns for both evolving and developed dunes. The cases listed in the key are presented in Tab. \ref{tab1}.}
	\label{fig:PDF_radial}
\end{figure}

\begin{figure}[ht]
	\includegraphics[width=0.95\linewidth]{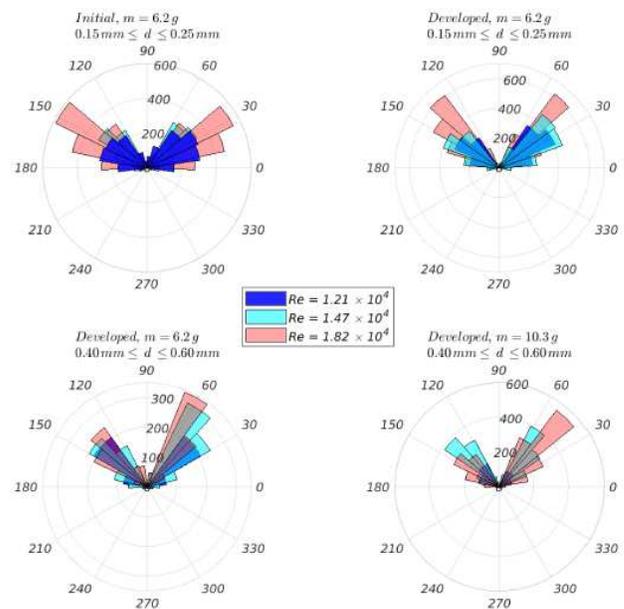}
	\caption{Frequencies of occurrence of the initial position of tracers migrating to the horns as a function of the angle with respect to the transverse direction. The water flow direction is 270$^{\circ}$. All cases listed in Tab. \ref{tab1} are shown.}
	\label{fig:polar}
\end{figure}

Figure \ref{fig:PDF_radial} shows the probability density functions of the initial position of grains migrating to horns as functions of the normalized radial position $|r_1 - r_c|/R$ and Fig. \ref{fig:polar} shows their frequencies of occurrence with respect to the transverse direction, for all tested conditions listed in Tab. \ref{tab1}, corresponding then to both evolving and developed barchans. For cases d to l, the centroid position is as shown in Fig. \ref{fig:dune_track}(d), while for cases a to c, the centroid is originally the center of the initial circle (horizontal projection of the initial pile), with small relative changes from the beginning to the end of the tests as the bedform evolves. The probabilities and frequencies for the larger grains over evolving barchans are not presented here and can be found in Ref. \cite{Alvarez3}. The PDFs were computed using a kernel smoothing function \cite{Bowman_Azzalini} in Fig. \ref{fig:PDF_radial} and the water flow direction is 270$^{\circ}$ in Fig. \ref{fig:polar}.

Regardless of the dune condition (evolving or developed), initial mass of the pile, grain size and water flow rate, much of the grains going to horns come from upstream regions on the periphery of the pile or dune, $|r_1 - r_c|/R$ $>$ 1 and 15$^{\circ}$ $\leq\,\phi\,\leq$ 70$^{\circ}$ and 120$^{\circ}$ $\leq\,\phi\,\leq$ 170$^{\circ}$. These results are in agreement with those for evolving barchans presented in Ref. \cite{Alvarez3}. We note that asymmetries in the plots of Fig. \ref{fig:polar} are due to dispersions in our experiments.

Figure \ref{fig:polar} shows that some of grains migrating to horns have their origin at the lateral flanks ($\phi\,\approx$ 0$^{\circ}$ and $\phi\,\approx$ 180$^{\circ}$), with smaller frequencies of occurrence, however, than those with origin at upstream positions. Another portion of the grains that were originally at the lateral flanks is carried away downstream from the dune by the water flow, as can be seen in the movies available as Supplemental Material \cite{Supplemental}. We note also in these movies a region consisting of a monolayer of grains on the periphery of the bedform, with the exceptions of the lee side and inner part of the horns. Because most of grains migrating to horns come from upstream regions on the periphery of the dune, many of them come thus from the monolayer region.

Based on the present data and those of Ref. \cite{Alvarez3}, we can conclude that, within the range of parameters of the present study, grains going to the horns of both evolving and developed subaqueous barchans describe circular paths while moving by rolling and sliding. This is different from the picture reported for the aeolian case, where bed load is characterized by salting grains that effectuate ballistic flights in the wind direction, impacting in many instances onto the dune surface \cite{Hersen_3, Schwammle, Andreotti_5}. In the aeolian case, horns would form from grains with original position at the lateral flanks of the initial pile, and thereafter would be maintained by grains coming from the lateral flanks of the developed barchan, but trajectories of grains and horns formation in the aeolian case rest to be investigated.

\subsection{\label{subsec:Res_C} Velocity distributions of grains migrating to horns}  

We computed the instantaneous transverse and streamwise velocities, $V_x$ and $V_y$, for all tracers that migrated to the horns. The velocities were computed by time differentiation of trajectories, based on algorithms described in Ref. \cite{Kelley}. The amount of moving particles varies with the water flow rate and the inverse of grain diameter; therefore, the statistics for $\theta$ = 0.14 take into account larger number of grains than those for smaller values of $\theta$. However, the velocity distributions that we obtained showed the same behavior for all tested conditions.

\begin{figure}[ht]	
	\begin{minipage}[c]{\columnwidth}
		\begin{center}
		\includegraphics[width=.65\linewidth]{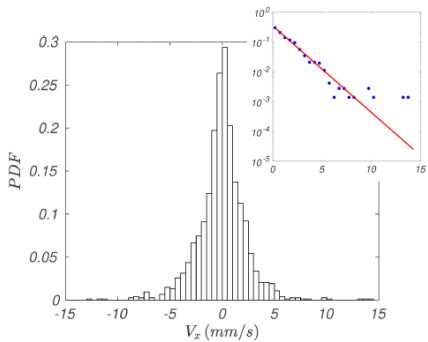}\\
			(a)
		\end{center}
	\end{minipage} \hfill
		\begin{minipage}[c]{\columnwidth}
    	\begin{center}
		\includegraphics[width=.65\linewidth]{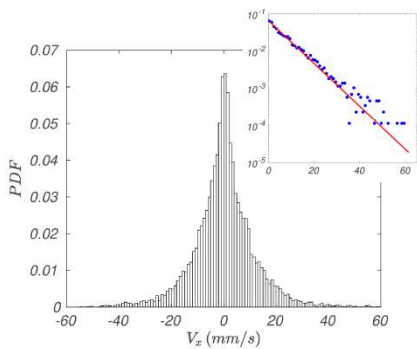}\\	
		(b)
		\end{center}
	\end{minipage}

	\caption{PDFs of transverse velocities $V_x$ of tracers migrating to horns. The insets are semi-log plots of PDFs for $V_x>0$, where the continuous line corresponds to a straight-line fit for this part of the distribution. Figures (a) and (b) correspond, respectively, to cases h and b listed in Tab. \ref{tab1}.} 
	\label{fig:PDF_Vx}
\end{figure}

Figures \ref{fig:PDF_Vx}(a) and \ref{fig:PDF_Vx}(b) present PDFs of the transverse velocities of tracers migrating to the barchan horns for cases h and b (Tab. \ref{tab1}), respectively. These PDFs are roughly symmetrical about $V_x$ = 0 and could be fitted by a Gaussian function. However, the distributions are too peaked to be adjusted in this manner and are better fitted by an exponential law. We note that the exponential fitting is rather poor for the larger velocities, as shown in the insets of Figs. \ref{fig:PDF_Vx}(a) and \ref{fig:PDF_Vx}(b), but the region of poor fitting corresponds to lower probabilities, of the order of 10$^{-3}$. The PDFs obtained for the other cases are similar.

\begin{figure}[ht]
		\begin{minipage}[c]{\columnwidth}
			\begin{center}
				\includegraphics[width=.65\linewidth]{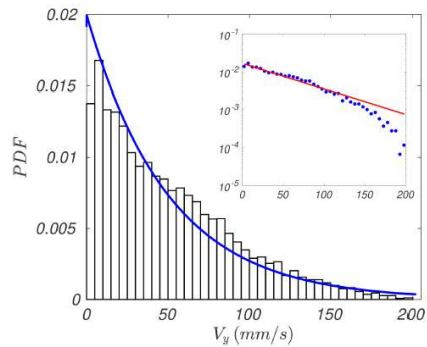}\\
				(a)
			\end{center}
		\end{minipage} \hfill
		\begin{minipage}[c]{\columnwidth}
			\begin{center}
				\includegraphics[width=.65\linewidth]{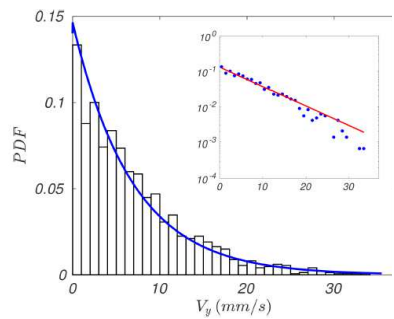}\\	
				(b)
			\end{center}
		\end{minipage}
	\caption{PDFs of streamwise velocities $V_y$ of tracers migrating to horns. The insets are semi-log plots of PDFs, where the continuous line corresponds to a straight-line fit. Figures (a) and (b) correspond, respectively, to cases h and b listed in Tab. \ref{tab1}.}
	\label{fig:PDF_Vy}
\end{figure}

The distributions of the streamwise velocities of the grains migrating to horns have some small negative values corresponding to grains trapped in the recirculation region, which affects some grains at the inner lateral flank of the horns (see Supplemental Material for movies showing the displacements of grains \cite{Supplemental}). Because the negative values have small magnitudes and very low probabilities, we did not consider them in our PDFs. Figures \ref{fig:PDF_Vy}(a) and \ref{fig:PDF_Vy}(b) present PDFs of $V_y$ of tracers migrating to the barchan horns for cases h and b (Tab. \ref{tab1}), respectively. As for $V_x$, the distributions of $V_y$ can be fitted by an exponential law, noting that fittings are rather poor for the larger velocities, as shown in the insets of Figs. \ref{fig:PDF_Vy}(a) and \ref{fig:PDF_Vy}(b). Again, as in the case for $V_x$, the region of poor fitting corresponds to lower probabilities, of the orders of 10$^{-4}$ and 10$^{-3}$, and the PDFs obtained for the other cases are similar.

We note that the PDFs of velocity distributions that we have obtained by considering only the grains going to horns resemble those obtained by Refs. \cite{Lajeunesse, Roseberry, Seizilles, Heyman} for bed load over flat beds, even if for the latter case the transverse displacement of grains is only of diffusive nature.

\subsection{\label{subsec:Res_D} Residence time of moving grains leaving the horns of developed barchans}

We investigate next the residence time of moving grains whose initial positions were on the horns of developed barchans. For that, we computed the time interval $t$ that each tracer took to leave the horns once they started moving as well as the traveled distance $\delta$. This concerns tracers that had stopped over the horns or that were previously buried and suddenly exposed by erosion. Therefore, we employ here the term \textit{residence time of moving grains} to make it clear that we are not dealing with the characteristic time of residence of grains within the horns, but with the time that, once moving, they take to leave the horns. This is different from the concept of residence time computed by Zhang et al. \cite{Zhang_D}, who investigated numerically the residence time of grains in the entire barchan before their ejection at the tips of horns. Zhang et al. \cite{Zhang_D} found that the residence time in the barchan is given by the surface of the longitudinal central slice of the dune divided by the input sand flux.

 \begin{figure}[th!]
	\includegraphics[width=0.60\linewidth]{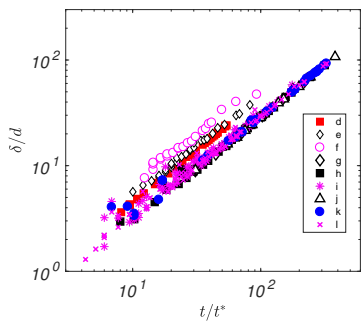}
	\caption{Traveled distance normalized by the grain diameter, $\delta /d$, as a function of the residence time of moving grains leaving the horns normalized by the settling time, $t/t^*$. A quasi-linear relationship is verified for all tested cases. The cases listed in the key are presented in Tab. \ref{tab1}.}
	\label{fig:dist_time}
\end{figure}

Figure \ref{fig:dist_time} presents $\delta$ normalized by the grain diameter as a function of $t$ normalized by a settling time $t^*$ defined as

\begin{equation}
t^* = \left(\frac{\rho \: d}{(\rho_s - \rho)g}\right)^{1/2},
\label{Eq:time}
\end{equation} 

\noindent in log-log scales, for all moving tracers with origin on the horns. Independently of the flow rate and grain size, $\delta /d$ varies quasi-linearly with $t/t^*$, grains under higher Shields numbers remaining shorter times on the horns (cases d, e and f); these grains, dragged by a stronger fluid flow, travel larger distances in shorter times. An exponential fit of the data presented in Fig. \ref{fig:dist_time} gives

\begin{equation}
\delta/d \sim (t/t^*)^{0.8702},
\label{Eq:time_fitting}
\end{equation} 

\noindent with $R^2$ = 0.997, i.e., a quasi-linear relation.

\begin{figure}
	\begin{minipage}[c]{\columnwidth}
		\begin{center}
			\includegraphics[width=.65\linewidth]{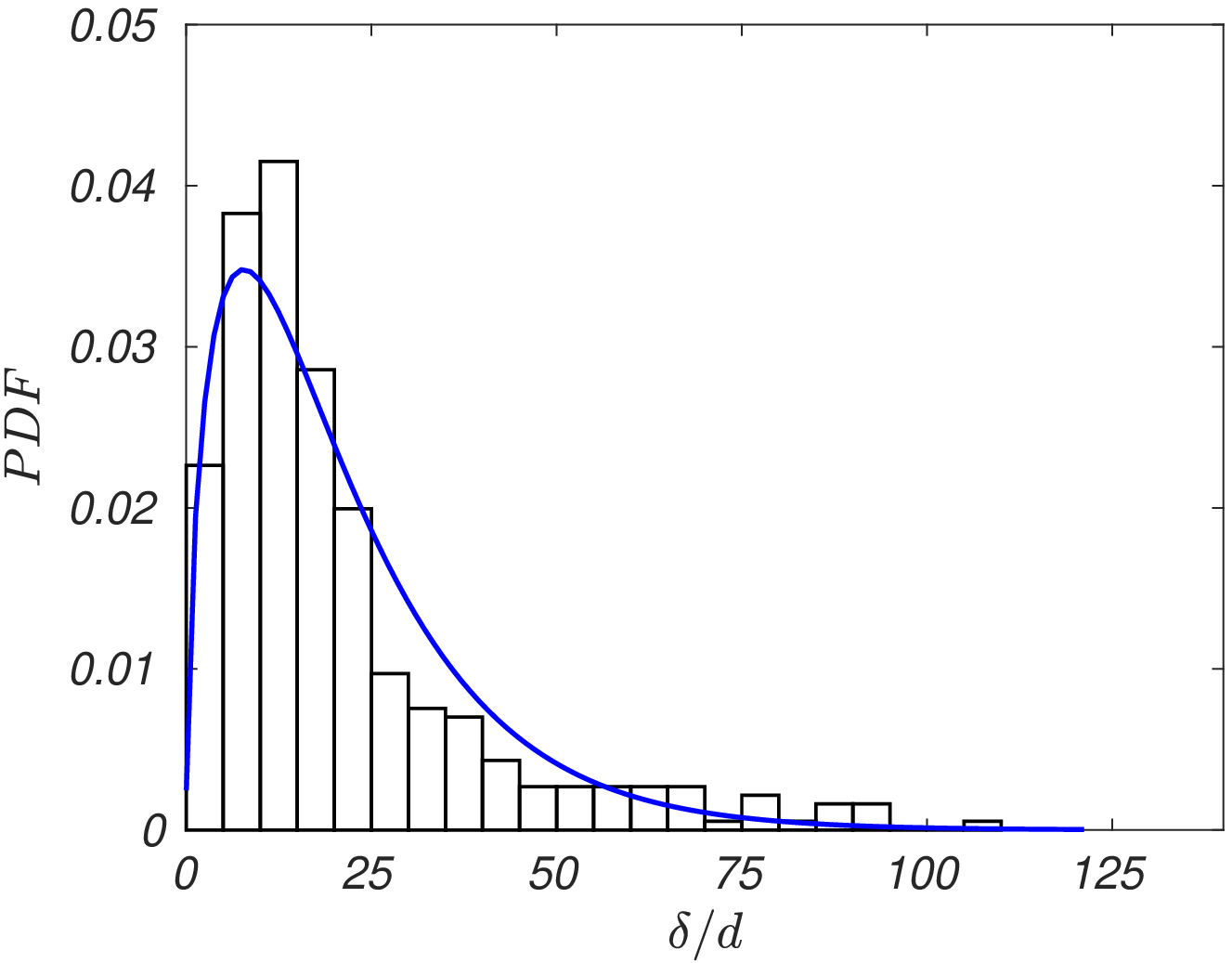}\\
			(a)
		\end{center}
	\end{minipage} \hfill
	\begin{minipage}[c]{\columnwidth}
		\begin{center}
			\includegraphics[width=.65\linewidth]{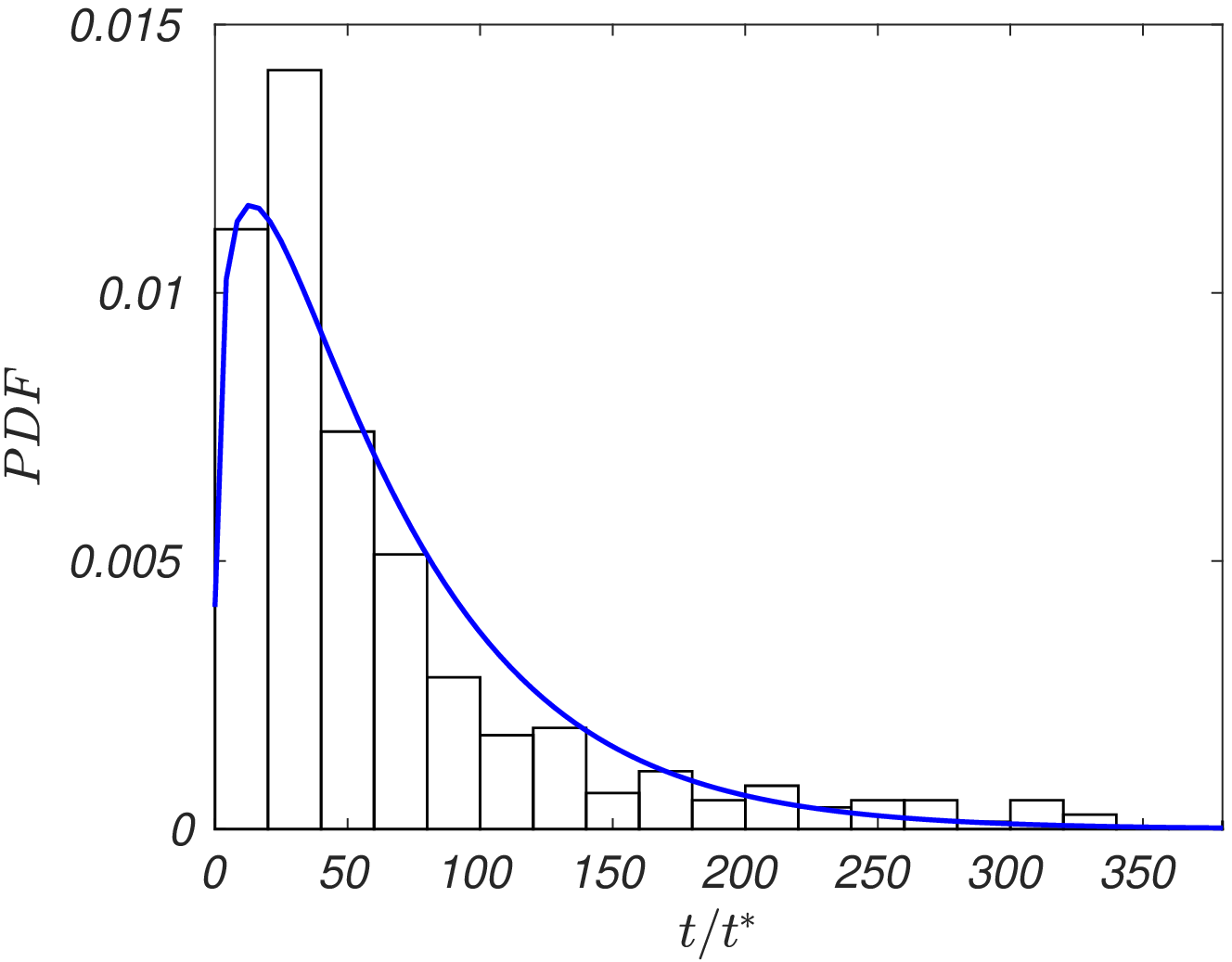}\\
			(b)
		\end{center}
	\end{minipage}
	\caption{PDFs of (a) traveled distance normalized by the grain diameter, $\delta /d$, and (b) residence time of moving grains leaving the horns normalized by the settling time, $t/t^*$. The continuous lines correspond to fittings using gamma functions. Cases d to l are included in these PDFs.}
	\label{fig:PDF_horn}
\end{figure}

In Alvarez and Franklin \cite{Alvarez}, we found that for developed barchans the horn length is $L_h \,\approx\, 10L_{drag}$. Therefore, in the case of subaqueous barchans $L_h \,\approx\, 25d$ and we expect that $\delta$ scales with this value. Figures \ref{fig:PDF_horn}(a) and \ref{fig:PDF_horn}(b) present the PDFs of the traveled distance normalized by the grain diameter, $\delta /d$, and residence time of moving grains normalized by the settling time, $t/t^*$, respectively, for cases d to l. The data presented in Figs. \ref{fig:PDF_horn}(a) and \ref{fig:PDF_horn}(b) were fitted by gamma functions, showed as continuous lines in these figures. For these functions, the most probable value of $\delta$ is 15$d$ (the mean value for the gamma fit being 20$d$) and the most probable residence time is 35$t^*$ (the mean value for the gamma fit being 65$t^*$). The most probable value found for $\delta$ is on the expected order of magnitude. Concerning the most probable residence time, it corresponds to dimensional values of 0.13 s $\leq\,t\,\leq$ 0.20 s; therefore, the characteristic velocity of grains leaving the horns is 0.015 m/s $\leq\,\delta / t\,\leq$ 0.057 m/s, which corresponds to 6 -- 16$\%$ of the cross-sectional mean velocity of the water. These values are in accordance with Lajeunesse et al. \cite{Lajeunesse}, Roseberry et al. \cite{Roseberry} and Penteado and Franklin \cite{Penteado}, although these works concerned plane beds.

\section{\label{sec:Conclu} CONCLUSIONS}

In this paper we presented experimental results on the dynamics of grains migrating to horns of both evolving and developed subaqueous barchans. Our results showed that the majority of these grains do not come from the lateral flanks of the initial heap or dune, as usually asserted in the aeolian case. Instead, we showed that in the subaqueous case most of grains migrating to horns come from upstream regions of the bedform, exhibiting significant transverse displacements. For these grains, irrespective of their size and strength of water flow, we found that the distributions of transverse and streamwise velocities are given by exponential functions, with the probability density functions of their magnitudes being similar to results obtained from previous studies on flat beds. We computed the residence time of moving grains, which we defined as the time taken by moving grains whose initial positions were on the horns of developed barchans to leave them, and the corresponding length. We found that the residence time and traveled distance are related following a quasi-linear relation, and that their most probable values are 35 times the settling time and 15 grain diameters, respectively. In addition, we showed that the characteristic velocity of these grains is of the same order of velocities reported for subaqueous bed load on plane beds. Our results change the way in which the crescentic shape of subaqueous barchans is explained. However, the physical mechanisms underlying the shape of barchan dunes, which we identified for the subaqueous case, cannot be precluded for bedforms found on terrestrial deserts and other planetary environments. Therefore, further numerical and experimental investigations on aeolian dunes are necessary to shed more light on the physical mechanisms leading to the formation of barchans in different environments.

\section{\label{sec:Ack} ACKNOWLEDGMENTS}

Carlos A. Alvarez is grateful to SENESCYT (Grant No. 2013-AR2Q2850) and to CNPq (Grant No. 140773/2016-9). Erick M. Franklin is grateful to FAPESP (Grants No. 2016/13474-9 and No. 2018/14981-7), to CNPq (Grant No. 400284/2016-2) and to FAEPEX/UNICAMP (Grants No. 2100/18 and No. 2112/19) for the financial support provided.

\bibliography{references}

\begin{thebibliography}{38}%
\makeatletter
\providecommand \@ifxundefined [1]{%
 \@ifx{#1\undefined}
}%
\providecommand \@ifnum [1]{%
 \ifnum #1\expandafter \@firstoftwo
 \else \expandafter \@secondoftwo
 \fi
}%
\providecommand \@ifx [1]{%
 \ifx #1\expandafter \@firstoftwo
 \else \expandafter \@secondoftwo
 \fi
}%
\providecommand \natexlab [1]{#1}%
\providecommand \enquote  [1]{``#1''}%
\providecommand \bibnamefont  [1]{#1}%
\providecommand \bibfnamefont [1]{#1}%
\providecommand \citenamefont [1]{#1}%
\providecommand \href@noop [0]{\@secondoftwo}%
\providecommand \href [0]{\begingroup \@sanitize@url \@href}%
\providecommand \@href[1]{\@@startlink{#1}\@@href}%
\providecommand \@@href[1]{\endgroup#1\@@endlink}%
\providecommand \@sanitize@url [0]{\catcode `\\12\catcode `\$12\catcode
  `\&12\catcode `\#12\catcode `\^12\catcode `\_12\catcode `\%12\relax}%
\providecommand \@@startlink[1]{}%
\providecommand \@@endlink[0]{}%
\providecommand \url  [0]{\begingroup\@sanitize@url \@url }%
\providecommand \@url [1]{\endgroup\@href {#1}{\urlprefix }}%
\providecommand \urlprefix  [0]{URL }%
\providecommand \Eprint [0]{\href }%
\providecommand \doibase [0]{http://dx.doi.org/}%
\providecommand \selectlanguage [0]{\@gobble}%
\providecommand \bibinfo  [0]{\@secondoftwo}%
\providecommand \bibfield  [0]{\@secondoftwo}%
\providecommand \translation [1]{[#1]}%
\providecommand \BibitemOpen [0]{}%
\providecommand \bibitemStop [0]{}%
\providecommand \bibitemNoStop [0]{.\EOS\space}%
\providecommand \EOS [0]{\spacefactor3000\relax}%
\providecommand \BibitemShut  [1]{\csname bibitem#1\endcsname}%
\let\auto@bib@innerbib\@empty
%</preamble>
\bibitem [{\citenamefont {Bagnold}(1941)}]{Bagnold_1}%
  \BibitemOpen
  \bibfield  {author} {\bibinfo {author} {\bibfnamefont {R.~A.}\ \bibnamefont
  {Bagnold}},\ }\href@noop {} {\emph {\bibinfo {title} {The Physics of Blown
  Sand and Desert Dunes}}}\ (\bibinfo  {publisher} {Chapman and Hall},\
  \bibinfo {address} {London},\ \bibinfo {year} {1941})\BibitemShut {NoStop}%
\bibitem [{\citenamefont {Hersen}\ \emph {et~al.}(2002)\citenamefont {Hersen},
  \citenamefont {Douady},\ and\ \citenamefont {Andreotti}}]{Hersen_1}%
  \BibitemOpen
  \bibfield  {author} {\bibinfo {author} {\bibfnamefont {P.}~\bibnamefont
  {Hersen}}, \bibinfo {author} {\bibfnamefont {S.}~\bibnamefont {Douady}}, \
  and\ \bibinfo {author} {\bibfnamefont {B.}~\bibnamefont {Andreotti}},\ }\href
  {\doibase 10.1103/PhysRevLett.89.264301} {\bibfield  {journal} {\bibinfo
  {journal} {Phys. Rev. Lett.}\ }\textbf {\bibinfo {volume} {89}},\ \bibinfo
  {pages} {264301} (\bibinfo {year} {2002})}\BibitemShut {NoStop}%
\bibitem [{\citenamefont {Herrmann}\ and\ \citenamefont
  {Sauermann}(2000)}]{Herrmann_Sauermann}%
  \BibitemOpen
  \bibfield  {author} {\bibinfo {author} {\bibfnamefont {H.~J.}\ \bibnamefont
  {Herrmann}}\ and\ \bibinfo {author} {\bibfnamefont {G.}~\bibnamefont
  {Sauermann}},\ }\href@noop {} {\bibfield  {journal} {\bibinfo  {journal}
  {Physica A (Amsterdam)}\ }\textbf {\bibinfo {volume} {283}},\ \bibinfo
  {pages} {24} (\bibinfo {year} {2000})}\BibitemShut {NoStop}%
\bibitem [{\citenamefont {Hersen}(2004)}]{Hersen_3}%
  \BibitemOpen
  \bibfield  {author} {\bibinfo {author} {\bibfnamefont {P.}~\bibnamefont
  {Hersen}},\ }\href@noop {} {\bibfield  {journal} {\bibinfo  {journal} {Eur.
  Phys. J. B}\ }\textbf {\bibinfo {volume} {37}},\ \bibinfo {pages} {507}
  (\bibinfo {year} {2004})}\BibitemShut {NoStop}%
\bibitem [{\citenamefont {Claudin}\ and\ \citenamefont
  {Andreotti}(2006)}]{Claudin_Andreotti}%
  \BibitemOpen
  \bibfield  {author} {\bibinfo {author} {\bibfnamefont {P.}~\bibnamefont
  {Claudin}}\ and\ \bibinfo {author} {\bibfnamefont {B.}~\bibnamefont
  {Andreotti}},\ }\href@noop {} {\bibfield  {journal} {\bibinfo  {journal}
  {Earth Plan. Sci. Lett.}\ }\textbf {\bibinfo {volume} {252}},\ \bibinfo
  {pages} {20} (\bibinfo {year} {2006})}\BibitemShut {NoStop}%
\bibitem [{\citenamefont {Parteli}\ and\ \citenamefont
  {Herrmann}(2007)}]{Parteli2}%
  \BibitemOpen
  \bibfield  {author} {\bibinfo {author} {\bibfnamefont {E.~J.~R.}\
  \bibnamefont {Parteli}}\ and\ \bibinfo {author} {\bibfnamefont {H.~J.}\
  \bibnamefont {Herrmann}},\ }\href {\doibase 10.1103/PhysRevE.76.041307}
  {\bibfield  {journal} {\bibinfo  {journal} {Phys. Rev. E}\ }\textbf {\bibinfo
  {volume} {76}},\ \bibinfo {pages} {041307} (\bibinfo {year}
  {2007})}\BibitemShut {NoStop}%
\bibitem [{\citenamefont {Sauermann}\ \emph {et~al.}(2001)\citenamefont
  {Sauermann}, \citenamefont {Kroy},\ and\ \citenamefont
  {Herrmann}}]{Sauermann_4}%
  \BibitemOpen
  \bibfield  {author} {\bibinfo {author} {\bibfnamefont {G.}~\bibnamefont
  {Sauermann}}, \bibinfo {author} {\bibfnamefont {K.}~\bibnamefont {Kroy}}, \
  and\ \bibinfo {author} {\bibfnamefont {H.~J.}\ \bibnamefont {Herrmann}},\
  }\href {\doibase 10.1103/PhysRevE.64.031305} {\bibfield  {journal} {\bibinfo
  {journal} {Phys. Rev. E}\ }\textbf {\bibinfo {volume} {64}},\ \bibinfo
  {pages} {031305} (\bibinfo {year} {2001})}\BibitemShut {NoStop}%
\bibitem [{\citenamefont {Andreotti}\ \emph
  {et~al.}(2002{\natexlab{a}})\citenamefont {Andreotti}, \citenamefont
  {Claudin},\ and\ \citenamefont {Douady}}]{Andreotti_1}%
  \BibitemOpen
  \bibfield  {author} {\bibinfo {author} {\bibfnamefont {B.}~\bibnamefont
  {Andreotti}}, \bibinfo {author} {\bibfnamefont {P.}~\bibnamefont {Claudin}},
  \ and\ \bibinfo {author} {\bibfnamefont {S.}~\bibnamefont {Douady}},\
  }\href@noop {} {\bibfield  {journal} {\bibinfo  {journal} {Eur. Phys. J. B}\
  }\textbf {\bibinfo {volume} {28}},\ \bibinfo {pages} {321} (\bibinfo {year}
  {2002}{\natexlab{a}})}\BibitemShut {NoStop}%
\bibitem [{\citenamefont {Andreotti}\ \emph
  {et~al.}(2002{\natexlab{b}})\citenamefont {Andreotti}, \citenamefont
  {Claudin},\ and\ \citenamefont {Douady}}]{Andreotti_2}%
  \BibitemOpen
  \bibfield  {author} {\bibinfo {author} {\bibfnamefont {B.}~\bibnamefont
  {Andreotti}}, \bibinfo {author} {\bibfnamefont {P.}~\bibnamefont {Claudin}},
  \ and\ \bibinfo {author} {\bibfnamefont {S.}~\bibnamefont {Douady}},\
  }\href@noop {} {\bibfield  {journal} {\bibinfo  {journal} {Eur. Phys. J. B}\
  }\textbf {\bibinfo {volume} {28}},\ \bibinfo {pages} {341} (\bibinfo {year}
  {2002}{\natexlab{b}})}\BibitemShut {NoStop}%
\bibitem [{\citenamefont {Kroy}\ \emph
  {et~al.}(2002{\natexlab{a}})\citenamefont {Kroy}, \citenamefont {Sauermann},\
  and\ \citenamefont {Herrmann}}]{Kroy_A}%
  \BibitemOpen
  \bibfield  {author} {\bibinfo {author} {\bibfnamefont {K.}~\bibnamefont
  {Kroy}}, \bibinfo {author} {\bibfnamefont {G.}~\bibnamefont {Sauermann}}, \
  and\ \bibinfo {author} {\bibfnamefont {H.~J.}\ \bibnamefont {Herrmann}},\
  }\href {\doibase 10.1103/PhysRevE.66.031302} {\bibfield  {journal} {\bibinfo
  {journal} {Phys. Rev. E}\ }\textbf {\bibinfo {volume} {66}},\ \bibinfo
  {pages} {031302} (\bibinfo {year} {2002}{\natexlab{a}})}\BibitemShut
  {NoStop}%
\bibitem [{\citenamefont {Kroy}\ \emph
  {et~al.}(2002{\natexlab{b}})\citenamefont {Kroy}, \citenamefont {Sauermann},\
  and\ \citenamefont {Herrmann}}]{Kroy_C}%
  \BibitemOpen
  \bibfield  {author} {\bibinfo {author} {\bibfnamefont {K.}~\bibnamefont
  {Kroy}}, \bibinfo {author} {\bibfnamefont {G.}~\bibnamefont {Sauermann}}, \
  and\ \bibinfo {author} {\bibfnamefont {H.~J.}\ \bibnamefont {Herrmann}},\
  }\href {\doibase 10.1103/PhysRevLett.88.054301} {\bibfield  {journal}
  {\bibinfo  {journal} {Phys. Rev. Lett.}\ }\textbf {\bibinfo {volume} {88}},\
  \bibinfo {pages} {054301} (\bibinfo {year} {2002}{\natexlab{b}})}\BibitemShut
  {NoStop}%
\bibitem [{\citenamefont {Kroy}\ \emph {et~al.}(2005)\citenamefont {Kroy},
  \citenamefont {Fischer},\ and\ \citenamefont {Obermayer}}]{Kroy_B}%
  \BibitemOpen
  \bibfield  {author} {\bibinfo {author} {\bibfnamefont {K.}~\bibnamefont
  {Kroy}}, \bibinfo {author} {\bibfnamefont {S.}~\bibnamefont {Fischer}}, \
  and\ \bibinfo {author} {\bibfnamefont {B.}~\bibnamefont {Obermayer}},\
  }\href@noop {} {\bibfield  {journal} {\bibinfo  {journal} {J. Phys. Condens.
  Matter}\ }\textbf {\bibinfo {volume} {17}},\ \bibinfo {pages} {S1229}
  (\bibinfo {year} {2005})}\BibitemShut {NoStop}%
\bibitem [{\citenamefont {Groh}\ \emph {et~al.}(2008)\citenamefont {Groh},
  \citenamefont {Wierschem}, \citenamefont {Aksel}, \citenamefont {Rehberg},\
  and\ \citenamefont {Kruelle}}]{Groh1}%
  \BibitemOpen
  \bibfield  {author} {\bibinfo {author} {\bibfnamefont {C.}~\bibnamefont
  {Groh}}, \bibinfo {author} {\bibfnamefont {A.}~\bibnamefont {Wierschem}},
  \bibinfo {author} {\bibfnamefont {N.}~\bibnamefont {Aksel}}, \bibinfo
  {author} {\bibfnamefont {I.}~\bibnamefont {Rehberg}}, \ and\ \bibinfo
  {author} {\bibfnamefont {C.~A.}\ \bibnamefont {Kruelle}},\ }\href {\doibase
  10.1103/PhysRevE.78.021304} {\bibfield  {journal} {\bibinfo  {journal} {Phys.
  Rev. E}\ }\textbf {\bibinfo {volume} {78}},\ \bibinfo {pages} {021304}
  (\bibinfo {year} {2008})}\BibitemShut {NoStop}%
\bibitem [{\citenamefont {Franklin}\ and\ \citenamefont
  {Charru}(2011)}]{Franklin_8}%
  \BibitemOpen
  \bibfield  {author} {\bibinfo {author} {\bibfnamefont {E.~M.}\ \bibnamefont
  {Franklin}}\ and\ \bibinfo {author} {\bibfnamefont {F.}~\bibnamefont
  {Charru}},\ }\href@noop {} {\bibfield  {journal} {\bibinfo  {journal} {J.
  Fluid Mech.}\ }\textbf {\bibinfo {volume} {675}},\ \bibinfo {pages} {199}
  (\bibinfo {year} {2011})}\BibitemShut {NoStop}%
\bibitem [{\citenamefont {L{\"a}mmel}\ \emph {et~al.}(2012)\citenamefont
  {L{\"a}mmel}, \citenamefont {Rings},\ and\ \citenamefont {Kroy}}]{Lammel}%
  \BibitemOpen
  \bibfield  {author} {\bibinfo {author} {\bibfnamefont {M.}~\bibnamefont
  {L{\"a}mmel}}, \bibinfo {author} {\bibfnamefont {D.}~\bibnamefont {Rings}}, \
  and\ \bibinfo {author} {\bibfnamefont {K.}~\bibnamefont {Kroy}},\ }\href@noop
  {} {\bibfield  {journal} {\bibinfo  {journal} {New J. Phys.}\ }\textbf
  {\bibinfo {volume} {14}},\ \bibinfo {pages} {093037} (\bibinfo {year}
  {2012})}\BibitemShut {NoStop}%
\bibitem [{\citenamefont {Parteli}\ \emph {et~al.}(2014)\citenamefont
  {Parteli}, \citenamefont {Dur{\'a}n}, \citenamefont {Bourke}, \citenamefont
  {Tsoar}, \citenamefont {P{\"o}schel},\ and\ \citenamefont
  {Herrmann}}]{Parteli4}%
  \BibitemOpen
  \bibfield  {author} {\bibinfo {author} {\bibfnamefont {E.~J.~R.}\
  \bibnamefont {Parteli}}, \bibinfo {author} {\bibfnamefont {O.}~\bibnamefont
  {Dur{\'a}n}}, \bibinfo {author} {\bibfnamefont {M.~C.}\ \bibnamefont
  {Bourke}}, \bibinfo {author} {\bibfnamefont {H.}~\bibnamefont {Tsoar}},
  \bibinfo {author} {\bibfnamefont {T.}~\bibnamefont {P{\"o}schel}}, \ and\
  \bibinfo {author} {\bibfnamefont {H.}~\bibnamefont {Herrmann}},\ }\href@noop
  {} {\bibfield  {journal} {\bibinfo  {journal} {Aeol. Res.}\ }\textbf
  {\bibinfo {volume} {12}},\ \bibinfo {pages} {121} (\bibinfo {year}
  {2014})}\BibitemShut {NoStop}%
\bibitem [{\citenamefont {Khosronejad}\ and\ \citenamefont
  {Sotiropoulos}(2017)}]{Khosronejad}%
  \BibitemOpen
  \bibfield  {author} {\bibinfo {author} {\bibfnamefont {A.}~\bibnamefont
  {Khosronejad}}\ and\ \bibinfo {author} {\bibfnamefont {F.}~\bibnamefont
  {Sotiropoulos}},\ }\href@noop {} {\bibfield  {journal} {\bibinfo  {journal}
  {J. Fluid Mech.}\ }\textbf {\bibinfo {volume} {815}},\ \bibinfo {pages} {117}
  (\bibinfo {year} {2017})}\BibitemShut {NoStop}%
\bibitem [{\citenamefont {Alvarez}\ and\ \citenamefont
  {Franklin}(2017{\natexlab{a}})}]{Alvarez}%
  \BibitemOpen
  \bibfield  {author} {\bibinfo {author} {\bibfnamefont {C.~A.}\ \bibnamefont
  {Alvarez}}\ and\ \bibinfo {author} {\bibfnamefont {E.~M.}\ \bibnamefont
  {Franklin}},\ }\href {\doibase 10.1103/PhysRevE.96.062906} {\bibfield
  {journal} {\bibinfo  {journal} {Phys. Rev. E}\ }\textbf {\bibinfo {volume}
  {96}},\ \bibinfo {pages} {062906} (\bibinfo {year}
  {2017}{\natexlab{a}})}\BibitemShut {NoStop}%
\bibitem [{\citenamefont {Alvarez}\ and\ \citenamefont
  {Franklin}(2018)}]{Alvarez3}%
  \BibitemOpen
  \bibfield  {author} {\bibinfo {author} {\bibfnamefont {C.~A.}\ \bibnamefont
  {Alvarez}}\ and\ \bibinfo {author} {\bibfnamefont {E.~M.}\ \bibnamefont
  {Franklin}},\ }\href {\doibase 10.1103/PhysRevLett.121.164503} {\bibfield
  {journal} {\bibinfo  {journal} {Phys. Rev. Lett.}\ }\textbf {\bibinfo
  {volume} {121}},\ \bibinfo {pages} {164503} (\bibinfo {year}
  {2018})}\BibitemShut {NoStop}%
\bibitem [{\citenamefont {Wang}\ and\ \citenamefont
  {Anderson}(2018)}]{Wang_Ch}%
  \BibitemOpen
  \bibfield  {author} {\bibinfo {author} {\bibfnamefont {C.}~\bibnamefont
  {Wang}}\ and\ \bibinfo {author} {\bibfnamefont {W.}~\bibnamefont
  {Anderson}},\ }\href {\doibase 10.1103/PhysRevE.98.033112} {\bibfield
  {journal} {\bibinfo  {journal} {Phys. Rev. E}\ }\textbf {\bibinfo {volume}
  {98}},\ \bibinfo {pages} {033112} (\bibinfo {year} {2018})}\BibitemShut
  {NoStop}%
\bibitem [{\citenamefont {Gadal}\ \emph {et~al.}(2019)\citenamefont {Gadal},
  \citenamefont {Narteau}, \citenamefont {du~Pont}, \citenamefont {Rozier},\
  and\ \citenamefont {Claudin}}]{Gadal}%
  \BibitemOpen
  \bibfield  {author} {\bibinfo {author} {\bibfnamefont {C.}~\bibnamefont
  {Gadal}}, \bibinfo {author} {\bibfnamefont {C.}~\bibnamefont {Narteau}},
  \bibinfo {author} {\bibfnamefont {S.~C.}\ \bibnamefont {du~Pont}}, \bibinfo
  {author} {\bibfnamefont {O.}~\bibnamefont {Rozier}}, \ and\ \bibinfo {author}
  {\bibfnamefont {P.}~\bibnamefont {Claudin}},\ }\href@noop {} {\bibfield
  {journal} {\bibinfo  {journal} {J. Fluid Mech.}\ }\textbf {\bibinfo {volume}
  {862}},\ \bibinfo {pages} {490} (\bibinfo {year} {2019})}\BibitemShut
  {NoStop}%
\bibitem [{\citenamefont {Engelund}(1970)}]{Engelund_1}%
  \BibitemOpen
  \bibfield  {author} {\bibinfo {author} {\bibfnamefont {F.}~\bibnamefont
  {Engelund}},\ }\href@noop {} {\bibfield  {journal} {\bibinfo  {journal} {J.
  Fluid Mech.}\ }\textbf {\bibinfo {volume} {42}},\ \bibinfo {pages} {225}
  (\bibinfo {year} {1970})}\BibitemShut {NoStop}%
\bibitem [{\citenamefont {Engelund}\ and\ \citenamefont
  {Fredsoe}(1982)}]{Engelund_Fredsoe}%
  \BibitemOpen
  \bibfield  {author} {\bibinfo {author} {\bibfnamefont {F.}~\bibnamefont
  {Engelund}}\ and\ \bibinfo {author} {\bibfnamefont {J.}~\bibnamefont
  {Fredsoe}},\ }\href@noop {} {\bibfield  {journal} {\bibinfo  {journal} {Ann.
  Rev. Fluid Mech.}\ }\textbf {\bibinfo {volume} {14}},\ \bibinfo {pages} {13}
  (\bibinfo {year} {1982})}\BibitemShut {NoStop}%
\bibitem [{\citenamefont {Elbelrhiti}\ \emph {et~al.}(2005)\citenamefont
  {Elbelrhiti}, \citenamefont {Claudin},\ and\ \citenamefont
  {Andreotti}}]{Elbelrhiti}%
  \BibitemOpen
  \bibfield  {author} {\bibinfo {author} {\bibfnamefont {H.}~\bibnamefont
  {Elbelrhiti}}, \bibinfo {author} {\bibfnamefont {P.}~\bibnamefont {Claudin}},
  \ and\ \bibinfo {author} {\bibfnamefont {B.}~\bibnamefont {Andreotti}},\
  }\href@noop {} {\bibfield  {journal} {\bibinfo  {journal} {Nature}\ }\textbf
  {\bibinfo {volume} {437}} (\bibinfo {year} {2005})}\BibitemShut {NoStop}%
\bibitem [{\citenamefont {Schw{\"a}mmle}\ and\ \citenamefont
  {Herrmann}(2005)}]{Schwammle}%
  \BibitemOpen
  \bibfield  {author} {\bibinfo {author} {\bibfnamefont {V.}~\bibnamefont
  {Schw{\"a}mmle}}\ and\ \bibinfo {author} {\bibfnamefont {H.~J.}\ \bibnamefont
  {Herrmann}},\ }\href@noop {} {\bibfield  {journal} {\bibinfo  {journal} {Eur.
  Phys. J. E}\ }\textbf {\bibinfo {volume} {16}},\ \bibinfo {pages} {57}
  (\bibinfo {year} {2005})}\BibitemShut {NoStop}%
\bibitem [{\citenamefont {Kidanemariam}\ and\ \citenamefont
  {Uhlmann}(2017)}]{Kidanemariam}%
  \BibitemOpen
  \bibfield  {author} {\bibinfo {author} {\bibfnamefont {A.~G.}\ \bibnamefont
  {Kidanemariam}}\ and\ \bibinfo {author} {\bibfnamefont {M.}~\bibnamefont
  {Uhlmann}},\ }\href@noop {} {\bibfield  {journal} {\bibinfo  {journal} {J.
  Fluid Mech.}\ }\textbf {\bibinfo {volume} {818}},\ \bibinfo {pages} {716}
  (\bibinfo {year} {2017})}\BibitemShut {NoStop}%
\bibitem [{\citenamefont {Seizilles}\ \emph {et~al.}(2014)\citenamefont
  {Seizilles}, \citenamefont {Lajeunesse}, \citenamefont {Devauchelle},\ and\
  \citenamefont {Bak}}]{Seizilles}%
  \BibitemOpen
  \bibfield  {author} {\bibinfo {author} {\bibfnamefont {G.}~\bibnamefont
  {Seizilles}}, \bibinfo {author} {\bibfnamefont {E.}~\bibnamefont
  {Lajeunesse}}, \bibinfo {author} {\bibfnamefont {O.}~\bibnamefont
  {Devauchelle}}, \ and\ \bibinfo {author} {\bibfnamefont {M.}~\bibnamefont
  {Bak}},\ }\href@noop {} {\bibfield  {journal} {\bibinfo  {journal} {Phys.
  Fluids}\ }\textbf {\bibinfo {volume} {26}},\ \bibinfo {pages} {013302}
  (\bibinfo {year} {2014})}\BibitemShut {NoStop}%
\bibitem [{\citenamefont {Lajeunesse}\ \emph {et~al.}(2010)\citenamefont
  {Lajeunesse}, \citenamefont {Malverti},\ and\ \citenamefont
  {Charru}}]{Lajeunesse}%
  \BibitemOpen
  \bibfield  {author} {\bibinfo {author} {\bibfnamefont {E.}~\bibnamefont
  {Lajeunesse}}, \bibinfo {author} {\bibfnamefont {L.}~\bibnamefont
  {Malverti}}, \ and\ \bibinfo {author} {\bibfnamefont {F.}~\bibnamefont
  {Charru}},\ }\href@noop {} {\bibfield  {journal} {\bibinfo  {journal} {J.
  Geophys. Res.}\ }\textbf {\bibinfo {volume} {115}},\ \bibinfo {pages}
  {F04001} (\bibinfo {year} {2010})}\BibitemShut {NoStop}%
\bibitem [{\citenamefont {Penteado}\ and\ \citenamefont
  {Franklin}(2016)}]{Penteado}%
  \BibitemOpen
  \bibfield  {author} {\bibinfo {author} {\bibfnamefont {M.~R.~M.}\
  \bibnamefont {Penteado}}\ and\ \bibinfo {author} {\bibfnamefont {E.~M.}\
  \bibnamefont {Franklin}},\ }\href {\doibase
  https://doi.org/10.1016/j.expthermflusci.2016.06.013} {\bibfield  {journal}
  {\bibinfo  {journal} {Exp. Therm. Fluid Sci.}\ }\textbf {\bibinfo {volume}
  {78}},\ \bibinfo {pages} {220 } (\bibinfo {year} {2016})}\BibitemShut
  {NoStop}%
\bibitem [{\citenamefont {Schlichting}(2000)}]{Schlichting_1}%
  \BibitemOpen
  \bibfield  {author} {\bibinfo {author} {\bibfnamefont {H.}~\bibnamefont
  {Schlichting}},\ }\href@noop {} {\emph {\bibinfo {title} {Boundary-Layer
  Theory}}}\ (\bibinfo  {publisher} {Springer},\ \bibinfo {address} {New
  York},\ \bibinfo {year} {2000})\BibitemShut {NoStop}%
\bibitem [{Sup()}]{Supplemental}%
  \BibitemOpen
  \href@noop {} {\bibinfo  {journal} {See Supplemental Material at [URL to be
  inserted by publisher] for the layout of the experimental device, some
  trajectories of grains superposed with a photograph of a barchan dune,
  additional graphics for the remaining experimental data, and movies showing
  the motion of grains over evolving and developed barchans}\ }\BibitemShut
  {NoStop}%
\bibitem [{\citenamefont {Alvarez}\ and\ \citenamefont
  {Franklin}(2017{\natexlab{b}})}]{Alvarez2}%
  \BibitemOpen
\bibfield  {journal} {  }\bibfield  {author} {\bibinfo {author} {\bibfnamefont
  {C.~A.}\ \bibnamefont {Alvarez}}\ and\ \bibinfo {author} {\bibfnamefont
  {E.~M.}\ \bibnamefont {Franklin}},\ }\href {\doibase
  https://doi.org/10.1016/j.physa.2016.08.071} {\bibfield  {journal} {\bibinfo
  {journal} {Physica A (Amsterdam)}\ }\textbf {\bibinfo {volume} {465}},\
  \bibinfo {pages} {725 } (\bibinfo {year} {2017}{\natexlab{b}})}\BibitemShut
  {NoStop}%
\bibitem [{\citenamefont {Kelley}\ and\ \citenamefont
  {Ouellette}(2011)}]{Kelley}%
  \BibitemOpen
  \bibfield  {author} {\bibinfo {author} {\bibfnamefont {D.~H.}\ \bibnamefont
  {Kelley}}\ and\ \bibinfo {author} {\bibfnamefont {N.~T.}\ \bibnamefont
  {Ouellette}},\ }\href@noop {} {\bibfield  {journal} {\bibinfo  {journal} {Am.
  J. Phys.}\ }\textbf {\bibinfo {volume} {79}},\ \bibinfo {pages} {267}
  (\bibinfo {year} {2011})}\BibitemShut {NoStop}%
\bibitem [{\citenamefont {Bowman}\ and\ \citenamefont
  {Azzalini}(1997)}]{Bowman_Azzalini}%
  \BibitemOpen
  \bibfield  {author} {\bibinfo {author} {\bibfnamefont {A.}~\bibnamefont
  {Bowman}}\ and\ \bibinfo {author} {\bibfnamefont {A.}~\bibnamefont
  {Azzalini}},\ }\href@noop {} {\emph {\bibinfo {title} {Applied Smoothing
  Techniques for Data Analysis}}}\ (\bibinfo  {publisher} {Oxford University
  Press},\ \bibinfo {address} {Oxford},\ \bibinfo {year} {1997})\BibitemShut
  {NoStop}%
\bibitem [{\citenamefont {Andreotti}\ \emph {et~al.}(2006)\citenamefont
  {Andreotti}, \citenamefont {Claudin},\ and\ \citenamefont
  {Pouliquen}}]{Andreotti_5}%
  \BibitemOpen
  \bibfield  {author} {\bibinfo {author} {\bibfnamefont {B.}~\bibnamefont
  {Andreotti}}, \bibinfo {author} {\bibfnamefont {P.}~\bibnamefont {Claudin}},
  \ and\ \bibinfo {author} {\bibfnamefont {O.}~\bibnamefont {Pouliquen}},\
  }\href {\doibase 10.1103/PhysRevLett.96.028001} {\bibfield  {journal}
  {\bibinfo  {journal} {Phys. Rev. Lett.}\ }\textbf {\bibinfo {volume} {96}},\
  \bibinfo {pages} {028001} (\bibinfo {year} {2006})}\BibitemShut {NoStop}%
\bibitem [{\citenamefont {Roseberry}\ \emph {et~al.}(2012)\citenamefont
  {Roseberry}, \citenamefont {Schmeeckle},\ and\ \citenamefont
  {Furbish}}]{Roseberry}%
  \BibitemOpen
  \bibfield  {author} {\bibinfo {author} {\bibfnamefont {J.~C.}\ \bibnamefont
  {Roseberry}}, \bibinfo {author} {\bibfnamefont {M.~W.}\ \bibnamefont
  {Schmeeckle}}, \ and\ \bibinfo {author} {\bibfnamefont {D.~J.}\ \bibnamefont
  {Furbish}},\ }\href@noop {} {\bibfield  {journal} {\bibinfo  {journal} {J.
  Geophys. Res.}\ }\textbf {\bibinfo {volume} {117}},\ \bibinfo {pages}
  {F03032} (\bibinfo {year} {2012})}\BibitemShut {NoStop}%
\bibitem [{\citenamefont {Heyman}\ \emph {et~al.}(2015)\citenamefont {Heyman},
  \citenamefont {Boh{\'o}rquez},\ and\ \citenamefont {Ancey}}]{Heyman}%
  \BibitemOpen
  \bibfield  {author} {\bibinfo {author} {\bibfnamefont {J.}~\bibnamefont
  {Heyman}}, \bibinfo {author} {\bibfnamefont {P.}~\bibnamefont
  {Boh{\'o}rquez}}, \ and\ \bibinfo {author} {\bibfnamefont {C.}~\bibnamefont
  {Ancey}},\ }\href@noop {} {\bibfield  {journal} {\bibinfo  {journal} {J.
  Geophys. Res.: Earth Surf.}\ }\textbf {\bibinfo {volume} {120}},\ \bibinfo
  {pages} {2529} (\bibinfo {year} {2015})}\BibitemShut {NoStop}%
\bibitem [{\citenamefont {Zhang}\ \emph {et~al.}(2014)\citenamefont {Zhang},
  \citenamefont {Yang}, \citenamefont {Rozier},\ and\ \citenamefont
  {Narteau}}]{Zhang_D}%
  \BibitemOpen
  \bibfield  {author} {\bibinfo {author} {\bibfnamefont {D.}~\bibnamefont
  {Zhang}}, \bibinfo {author} {\bibfnamefont {X.}~\bibnamefont {Yang}},
  \bibinfo {author} {\bibfnamefont {O.}~\bibnamefont {Rozier}}, \ and\ \bibinfo
  {author} {\bibfnamefont {C.}~\bibnamefont {Narteau}},\ }\href@noop {}
  {\bibfield  {journal} {\bibinfo  {journal} {J. Geophys. Res.: Earth Surf.}\
  }\textbf {\bibinfo {volume} {119}},\ \bibinfo {pages} {451} (\bibinfo {year}
  {2014})}\BibitemShut {NoStop}%
\end{thebibliography}%

\end{document}